\documentclass[12]{article}
\usepackage{graphics,graphicx,amsbsy,amssymb}

\newcommand{\UU}{{\boldmath \mbox{$U$}}}
\newcommand{\JJ}{{\boldmath \mbox{$J$}}}
\newcommand{\YY}{{\boldmath \mbox{$Y$}}}

\newcommand{\uu}{{\boldmath \mbox{$u$}}}

\newcommand{\ww}{{\boldmath \mbox{$w$}}}

\newcommand{\WW}{{\boldmath \mbox{$W$}}}
\newcommand{\XX}{{\boldmath \mbox{$X$}}}

\newcommand{\rr}{{\boldmath \mbox{$r$}}}

\newcommand{\vv}{{\boldmath \mbox{$v$}}}

\newcommand{\nn}{{\boldmath \mbox{$n$}}}

\newlength{\defbaselineskip}
\newcommand{\setlinespacing}[1]%
           {\setlength{\baselineskip}{#1 \defbaselineskip}}

\title{\textbf{
Multiscale Thermodynamics }}
\author{Miroslav Grmela \footnote{e-mail:miroslav.grmela@polymtl.ca}
\vspace{0.9cm}\\
 \'{E}cole Polytechnique de Montr\'{e}al,
  C.P.6079 suc. Centre-ville,\\
 Montr\'{e}al, H3C 3A7,  Qu\'{e}bec, Canada}

 \date{}

\begin{document}

\maketitle

\pagenumbering{roman}
\tableofcontents
\newpage
\pagenumbering{arabic}

\begin{abstract}

Multiscale thermodynamics is a theory of relations among  levels of investigation of complex systems. It includes the classical
equilibrium thermodynamics
as a special case but it is applicable to both static and time evolving processes in externally and internally driven macroscopic systems that are
far from equilibrium and are investigated on  microscopic, mesoscopic, and macroscopic levels.
In this paper we formulate the multiscale thermodynamics, explain its origin,  and illustrate  it  in   mesoscopic dynamics that combines  levels.

\end{abstract}

\section{Introduction}

\textit{A level of investigation} is a collection of results of certain type of experimental observations (different for different levels) made on
complex systems together with a theory that allows to organize them, to reproduce them, and to make predictions. The theory, based on the
insight inspired by the experimental data and by investigating  relations to nearby levels involving less or more details,  offers also an
understanding of the physics involved.
For instance, the \textit{equilibrium level} with the energy $E$, number of moles $N$, and volume $V$ serving as state variables  \cite{*Gibbs} and  the
\textit{microscopic level} with position and momenta of $\sim 10^{23}$ particles composing the macroscopic system serving as state variable
are examples  of two different autonomous levels of description. The latter is more microscopic (it takes into account more details) than
the former.
We  call the latter level an \textit{upper level} and the former a \textit{lower level}.

\textit{Multiscale thermodynamics is a theory of relations among different levels. }

Hamilton's  mechanics, classical thermodynamics, fluid mechanics, Boltzmann's kinetic theory,  Gibbs' equilibrium statistical mechanics,
and extensive studies  of relations among them
provide methods, tools, and also an inspiration to formulate a multiscale thermodynamics  of which all these classical investigations are particular
realizations. The multiscale thermodynamics provides a framework
for investigating static and dynamic aspects of reductions from an upper  to
a lower level with no constrains to  the closeness to equilibrium or to  the absence of
external or internal  driving forces.

Our objective in this paper is to formulate  the  multiscale thermodynamics as a passage \textit{upper level} $\rightarrow$ \textit{lower level},
(in Sections \ref{structure} and \ref{rt}), to present classical investigations of mesoscopic dynamics through  the eyes of the multiscale
thermodynamics (in Section \ref{realizations}), and to demonstrate its application in the mesoscopic dynamics in which levels are combined
(in Section \ref{hierarchy}).

\section{Structures in Multiscale Thermodynamics}\label{structure}

Let $\mathfrak{L}$, $\mathcal{L}$ and \textit{l} be three autonomous levels. The level $\mathfrak{L}$ involves more details than the level
$\mathcal{L}$ that in turn involves more details than the level \textit{l}. We shall call the levels involving more details
\textit{upper levels} or also more microscopic levels, the levels involving less details are called \textit{lower levels} or also more
macroscopic levels. We investigate the chain
\begin{equation}\label{chain}
\longrightarrow\mathfrak{L}\longrightarrow \mathcal{L}\longrightarrow l\longrightarrow
\end{equation}
where $\longrightarrow$  represents \textit{reduction} in which unimportant details are ignored and important overall features emerge.
In the diagram (\ref{chain}), the way up (i.e. towards more microscopic levels) is to the left and the way down (i.e. towards  more
macroscopic levels)  is to the right.
The level $\mathcal{L}$ in the diagram
has two structures, one is \textit{reduced structure} arising in the reduction $\mathfrak{L}\longrightarrow \mathcal{L}$ and the other
is the \textit{reducing structure} arising in the reduction $\mathcal{L}\longrightarrow l$. Every structure, both reduced and reducing,
consists of a \textit{thermodynamic relation} and a \textit{vector field}. The former
generates the geometry and the latter the time evolution. Both depend on the other  level that is involved in the reduction (i.e. on the level
$\mathfrak{L}$
 in the case of the reduced structure and on the level \textit{l} in the case of the reducing structure).
Every mesoscopic level $\mathcal{L}$ that has neighbours on both the left and the right sides in the chain (\ref{chain}) has thus   reduced
and a reducing thermodynamic relations and  reduced and reducing vector fields. In general, all the reduced  structures
will depend on the choice of the level on its left side (i.e. the level from which it is reduced)
and the reducing structures on the choice of the level on its right side (i.e. the level to which it is reducing).

The passages \textit{upper level} $\longrightarrow$ \textit{lower level} representing the reduction process can be mathematically formulated in two
ways, one called a \textit{time-evolution  passage} and the other a maximum entropy passage (in short, \textit{MaxEnt passage}). The former is
a mathematical formulation of the time evolution process that prepares the macroscopic systems under investigation for experimental observations
on the lower level. The latter is a map transforming  initial states  on the level $\mathcal{L}$ into the final states  by following
the preparation process to its conclusion.  In other words, the latter is a property of solutions of the time evolution equations introduced
in the former.

If the focus of the investigation of the relations between the levels $\mathcal{L}$ and \textit{l} is put on the rates of the processes involved
rather than on the processes themselves then the resulting passages and structures form what we call \textit{multiscale rate thermodynamics}. The
two passages: time-evolution passage and MaxEnt passage become in the rate thermodynamics \textit{rate time-evolution passage} and maximum
rate-entropy passage that we write as \textit{MaxRent passage}. The structures become reducing and reduced rate structures.

Since we shall be investigating  in this paper  also direct links  $\mathfrak{L}\longrightarrow l$ and we shall be comparing them
with the composed links  $ \mathfrak{L}\longrightarrow \mathcal{L}\longrightarrow l $,  it is more convenient to replace the chain (\ref{chain})
with an oriented graph in which the levels $\mathfrak{L}, \mathcal{L},l,... $  are vertices and the reductions are links
connecting them. The links are directed from upper to lower levels (see more in
Section \ref{statmech}).
Altogether, the level $\mathcal{L}$ in the graph is equipped with many structures depending on the levels with which it is
compared (with which it is linked). The reduction represented by $\longrightarrow$ has two versions: time-evolution and MaxEnt. Moreover,
if the vector fields rather than
state spaces are compared then  the reducing and reduced structures become  reducing and reduced  rate structures and the total number
of structures doubles.
All the structures are not however independent
We shall see some of the dependencies  below in this paper.

Before proceeding to the actual formulation of the structure and the passages we
emphasize that the term "reduction" has  in this paper the same meaning as "emergence". Some details on the upper level are lost in the
reduction from  an upper level to a lower level but at the same time an emerging overall pattern is gained. The process of reduction, as well
as the processes conducive to an emergence of overall features (pattern-recognition processes), involve both a loss and a gain. The lower level
is inferior to the upper level in the amount of details but superior in the ability to display overall patterns.

\subsection{Time-evolution Passage }\label{te}

We begin by formulating  the  reducing structure on a level $\mathcal{L}$  that is being compared with  a lower level \textit{l}.
Both  levels $\mathcal{L}$ and \textit{l} are assumed to be well established and autonomous. This means that the macroscopic systems
whose behavior are found
to be well described on both levels can be prepared for the level \textit{l}. The time evolution describing the preparation process is the
reducing time evolution taking place on $\mathcal{L}$. For example, if the level \textit{l} is the equilibrium level, the preparation process
consists of leaving the macroscopic systems free of external influences  and internal constraints sufficiently long time (see more in
Section \ref{Boltzmann}).

Investigations of many  pairs of levels ($\mathcal{L}$,\textit{l}) (see more in Section \ref{realizations}) revealed the following structure of
the reducing time evolution. Let $x$
denote the state variable (for instance $x$ is the one particle distribution function in kinetic theory) used on the level $\mathcal{L}$.
 The vector field generating the
reducing time evolution on the level $\mathcal{L}$ is a sum of two terms, one is the Hamiltonian vector field and the other gradient vector field.
The former is an
inheritance of the mechanics seen on the microscopic level and the latter drives trajectories (i.e. solutions of the governing equations)
towards the time evolution on the level \textit{l}.
Both the Hamiltonian and the gradient parts of the vector fields are  gradients of a
potential (i.e. co-vectors) transformed into vectors by a geometrical structure. In the Hamiltonian part the potential is the energy and the
geometrical structure the Poisson structure (in the simplest case a skewsymmetric matrix). In the gradient part the potential is the entropy
and the geometrical structure is the metric structure (in the simplest case a symmetric matrix). Both geometrical structures are degenerate in
order to guarantee the conservation of energy and the increase of entropy. Comments  concerning the provenance of the reducing time evolution
are in Section \ref{Boltzmann}.

We now proceed to the mathematical formulation. The quantities characterizing states  are denoted by $x$ on the upper level and $y$ on the lower level.
All other  quantities belonging to the upper level are denoted with the upper index $\uparrow$ and to  the lower level with the upper index
$\downarrow$. The state space on the upper level is denoted $M^{\uparrow}$ (i.e. $x\in M^{\uparrow}$ ) and the state space on the lower level
$M^{\downarrow}$ (i.e. $y\in M^{\downarrow}$.
A special notation is used for the equilibrium level;
the state variables are $(E,N)$, where $E$ is the energy per unit volume and $N$ the number of particles per unit volume, the state space is
$M^{(eq)}$ (i.e. $(E,N)\in M^{(eq)}$). We use a shorthand notation  for derivatives: $A_x=\frac{\partial A}{\partial x}$,
where $A:M^{\uparrow} \rightarrow \mathbb{R}$ and $\frac{\partial}{\partial x}$ is an appropriate functional derivative in the
case when $M^{\uparrow}$ is an infinite dimensional space.

We begin  the mathematical formulation of the reducing structure on the level $\mathcal{L}$ with the equilibrium level playing the role of
the level \textit{l} with which we are comparing the level  $\mathcal{L}$. First, we need a map
\begin{equation}\label{upeq}
M^{\uparrow}\rightarrow M^{(eq)};\,\,x\mapsto (E^{\uparrow}(x),N^{\uparrow}(x))
\end{equation}
The time evolution taking place in the process of preparing the macroscopic system under investigation for the equilibrium level (the reducing
time evolution) brings $x\in M^{\uparrow}$ to $\mathcal{M}^{\uparrow (eq)}\subset M^{\uparrow}$ that is in one-to-one relation to the equilibrium
state space $M^{(eq)}$. Our goal now is to identify the reducing time evolution. First, we turn to the Hamiltonian part, then to the gradient part,
and finally we combine them.

\subsubsection{Hamiltonian time evolution}\label{HHam}

The Hamiltonian part of the time evolution is governed by \cite{Arnoldmech}
\begin{equation}\label{Ham1}
\dot{x}=L^{\uparrow}E^{\uparrow}_x
\end{equation}
The operator $L^{\uparrow}$ is a Poisson bivector which means that the bracket defined by
\begin{equation}\label{PB}
\{A,B\}^{\uparrow}=<A_x,L^{\uparrow}B_x>
\end{equation}
is a Poisson bracket (i.e. $\{A,B\}^{\uparrow}=-\{B,A\}^{\uparrow}$ and the Jacobi identity
\\ $\{A,\{B,C\}^{\uparrow}\}^{\uparrow}+\{B,\{C,A\}^{\uparrow}\}^{\uparrow}+\{C,\{A,B\}^{\uparrow}\}^{\uparrow}=0$ holds), $A,B,C$ are
sufficiently regular real valued functions of $x\in M^{\uparrow}$ and $<,>$ denote the pairing in the space $M^{\uparrow}$.
From the physical point of view, the bivector $L^{\uparrow}$ expresses mathematically the kinematics of the chosen state variable
$x\in M^{\uparrow}$. For example, if $x=(\rr,\vv)$, where $\rr$ is the position coordinate and $\vv$ the momentum of one particle,
then $L^{\uparrow}=\left(\begin{array}{cc}0&1\\-1&0\end{array}\right)$ expressing mathematically the cotanget budle structure of
$M^{\uparrow}$. Other examples are in Section \ref{Boltzmann}

We note that the energy $E^{\uparrow}(x)$ is conserved in the time evolution governed by (\ref{Ham1}) since
$\dot{E^{\uparrow}}=\{E^{\uparrow},E^{\uparrow}\}=0$. In order to conserve other potentials in the time evolution (\ref{Ham1}),
the Poisson bivector $L^{\uparrow}$ has to be degenerate. We say that $C^{\uparrow}(x)$ is a \textit{Casimir of the Poisson
bracket} $\{A,B\}^{\uparrow}$ if $\{A,C\}^{\uparrow}=0 \,\,\,\forall A$. Consequently,  $\dot{C^{\uparrow}}=\{C^{\uparrow},E^{\uparrow}\}=0$
We require that the Poisson bivector $L^{\uparrow}$ arising in the
Hamiltonian part (\ref{Ham1}) of the reducing time evolution is degenerate with the number of moles $N^{\uparrow}(x)$ in (\ref{upeq}) and the
entropy $S^{\uparrow}(x)$ introduced below in the gradient part of the reducing time evolution are its Casimirs.

\subsubsection{Gradient time evolution}\label{grevol}

The Hamiltonian dynamics (\ref{Ham1}) can be transformed into a reducing dynamics by making the following three-step reduction:
(Step 1) All trajectories are found (i.e. all solutions of (\ref{Ham1}) passing through all $x\in M^{\uparrow}$  for an ensamble  of
$E^{\uparrow}(x)$ is found. The collection of all such trajectories is called a \textit{phase portrait}. (Step 2) A pattern is extracted in the phase
portrait. (Step 3) The  pattern is interpreted  as a phase portrait of the dynamics on the lower level \textit{l}. Let us assume that the above
three steps have been made. The result is expressed in the reducing time evolution. By following it to its conclusion we arrive on
the level \textit{l} (i.e. in the context of this section, on the equilibrium level). The equation governing the reducing time
evolution is  (\ref{Ham1}) modified by including
a seed of dissipation. The dissipation makes to disappear
unimportant details (this is the  loss in the reduction) and makes to emerge the  pattern  (this is the gain in the reduction). How to formulate the
dissipation?

The most significant contribution of the classical thermodynamics  is the MaxEnt principle (see more in Section \ref{realizations}). The pattern is
 revealed and  unimportant details discarded by maximizing a new potential $S^{\uparrow}(x)$ called a reducing entropy. The role of the new
 potential  $S^{\uparrow}(x)$ in mechanics is to reveal some overall features  of solutions to its governing equations. It is thus a potential
 that feels already some overall features of solutions and feeds them back to the initial upper vector field (see more in Section \ref{Boltzmann}).

When reaching the lower level, the reducing entropy becomes, on the lower level, the reduced entropy $S^{\downarrow}(y)$,
where $y$ is the state variable  used on the lower level. In particular, if the lower level is the equilibrium level then $S^{\downarrow}(E,N)$ is
the classical equilibrium entropy.

The simplest time evolution making the entropy $S^{\uparrow}(x)$ to grow  is the gradient dynamics \cite{grad1}, \cite{grad2}
\begin{equation}\label{gr1}
\dot{x}=\Lambda^{\uparrow}S^{\uparrow}_x
\end{equation}
where $\Lambda^{\uparrow}$ is the positive definite operator. Indeed, (\ref{gr1}) implies
\begin{equation}\label{gr222}
\dot{S}^{\uparrow}=<S^{\uparrow}_x\Lambda^{\uparrow}S^{\uparrow}_x>>0
\end{equation}
where $<,>$ denotes the pairing in $M^{\uparrow}$.

A straightforward generalization of (\ref{gr1}) is a \textit{dissipation-potential gradient dynamics} (see more in Section \ref{chemkin})
\begin{equation}\label{gr2}
\dot{x}=\left[\Xi^{\uparrow}_{x^*}(x;X)\right]_{x^*=S^{\uparrow}_x}
\end{equation}
where $\Xi^{\uparrow}$, called a \textit{dissipation potential} \cite{*gengrad},  is a real valued function of $(x,X)$ such that:
\begin{eqnarray}\label{disspot}
&&(i)\,\, \Xi^{\uparrow}(x,0)=0 \nonumber\\
&&(ii)\, \Xi^{\uparrow}\,\, reaches\,\, its\,\, minimum\,\, at\,\, X=0 \nonumber \\
&&(iii)\, \Xi^{\uparrow}\,\,  is\,\, a\,\,convex\,\, function\,\,of\,\,X\,\,in\,\, a\,\, neighborhood\,\, of\,\, X=0\nonumber \\
&&(iv)\,  X\,\,is\,\, a\,\, linear\,\, function\,\, of\,\, x^*\,\,such\,\, that \nonumber \\
&&\,\,\,\,\,\,\,\,\,\,<x^*,\Xi_{x^*}>=a<X,\Xi_{X}>,\,\, where\,\, a>0
\end{eqnarray}
We note that the potential that generates the time evolution (\ref{gr2}) is the entropy $S(x)$. The dissipation potential
is a different type of potential. It does not generate the time evolution but it plays the role of the
geometrical structure transforming the gradient $S_x(x)$ of the generating potential (that is a co-vector) into a vector field.
 The right hand side of (\ref{gr2}) becomes the same as the right hand side of (\ref{gr1}) when $X=x^*$ and
 $\Xi^{\uparrow}=\frac{1}{2}<X,\Lambda^{\uparrow}X>$. Regarding the requirement (iv) in (\ref{disspot}), we note that it is, of course,
 satisfied for $X=x^*$. In the case of $x$ being a field (i.e. a function of the position coordinate $\rr$) then the property (iv) is for
 example satisfied for $X=\nabla x^*$ provided the boundary condition guarantee the disappearance of integrals over the boundary.
 An example illustrating the requirement (iv) in the case when ($x=$ \textit{one particle distribution function})  is presented in Section \ref{Boltzmann}
 in (\ref{Bpot}).

A real valued function $\mathfrak{C}^{\uparrow }(x)$ for which
$X(\mathfrak{C}^{\uparrow }_x)=0$
are called \textit{gradient Casimirs}. They
 are  conserved (due to the property (iv) in (\ref{disspot})) in the dissipation-potential gradient time evolution (\ref{gr2}).

The inequality (\ref{gr222})  becomes
\begin{equation}\label{212}
\dot{S}^{\uparrow}=<S^{\uparrow}_x\left[\Xi^{\uparrow}_{x^*}(x;x^*)\right]_{x^*=S^{\uparrow}_x}>
=a\left[<X^*,\Xi_{X^*}>\right]_{x^*=S^{\uparrow}_x}>\geq 0
\end{equation}
where the equality holds for the \textit{dissipation equilibrium states}
\begin{equation}\label{diss1}
\mathcal{M}^{\uparrow (deq)}=\{x\in M^{\uparrow}|\Phi^{\uparrow (diss)}_x=0\}
\end{equation}
The upper thermodynamic potential $\Phi^{\uparrow }$ is given by
\begin{equation}\label{diss2}
\Phi^{\uparrow}(x;\mathfrak{C}^{*})=-S^{\uparrow}(x)+<\mathfrak{C}^{*},\mathfrak{C}^{\uparrow}(x)>
\end{equation}
and $\mathfrak{C}^{*}$ can be seen as being Lagrange multipliers since
(\ref{diss1}) can be read as maximization of the entropy  $S^{\uparrow}(x)$ subjected to
constraints $\mathfrak{C}^{\uparrow }(x)$.

If the thermodynamic potential $\Phi^{\uparrow }(x;\mathfrak{C}^{*})$ is convex then the inequality (\ref{212}) makes it possible
to consider $\Phi^{\uparrow}(x;\mathfrak{C}^{*})$ as a Lyapunov function for the approach (as $t\rightarrow \infty$) of solutions
to (\ref{gr2}) to $\mathcal{M}^{\uparrow (deq)}$.

The size of the manifold $\mathcal{M}^{\uparrow (deq)}$ makes it also possible to give a meaning to the strength of dissipation. We say that the
dissipation generated by a dissipation potential $\Xi^{(1)}$ is weaker than the dissipation generated
by $\Xi^{(2)}$ if $\mathcal{M}^{(deq)1}\supset \mathcal{M}^{(deq)2}$. The weakest dissipation is, of course, no dissipation
when $\mathcal{M}^{(deq)}\equiv M^{\uparrow}$.

\subsubsection{GENERIC time evolution}\label{GENsection}

We now combine the seed of dissipation (\ref{gr2}) with the Hamiltonian time evolution (\ref{Ham1}) in a way that
 essential features of mechanics (in particular the energy conservation) and also  of the gradient
dynamics (in particular the growth of the entropy) are preserved. Both from the physical and the mathematical point of view,
the combination can be best argued in the
setting of the contact geometry  that we present in Section \ref{CS} below. Here we introduce the combined dynamics in the form
\begin{equation}\label{GEN}
\dot{x}=L^{\uparrow}E^{\uparrow}_x+\left[\Xi^{\uparrow}_{x^*}\right]_{x^*=S^{\uparrow}_x}
\end{equation}
that is called GENERIC (its provenance is recalled in Section \ref{realizations}). Solutions to (\ref{GEN}) are required
to satisfy  the following properties:
\begin{eqnarray}\label{gen1}
\dot{E}^{\uparrow}&=&0\nonumber \\
\dot{N}^{\uparrow}&=&0\nonumber \\
\dot{S}^{\uparrow}&\geq &0
\end{eqnarray}
The first two conservations (conservations of the energy and of the number of moles) are dictated by mechanics. In (\ref{GEN}) we are modifying  the
Hamiltonian mechanics (\ref{Ham1}) by adding  dissipation but the essence of  mechanics must remain intact.
The modification is  made in order  to bring to  light overall features of  solutions to (\ref{Ham1}).
The modified Hamilton  equation (\ref{GEN}) represents still mechanics.  The total energy and the total mass conservations are essential to mechanics.
The last inequality   in (\ref{gen1}) (the entropy inequality) is a new feature, brought about by the modification,
that is fundamental for revealing the overall features (for proving that solutions to (\ref{GEN}) approach
equilibrium states).

Before proceeding to the proof, we note that (\ref{gen1}) are guaranteed if both the Poisson and the gradient structures are degenerate in
the sense that
\begin{eqnarray}\label{degen}
&& N^{\uparrow}(x), S^{\uparrow}(x)\,\, are \,\,Casimirs \nonumber \\
&&N^{\uparrow}(x), E^{\uparrow}(x)\,\, are\,\, gradient\,\, Casimirs
\end{eqnarray}

The proof of  the approach to equilibrium begins with  introducing  an upper reducing thermodynamic potential
\begin{equation}\label{eq4}
\Phi^{\uparrow}(x;E^*,N^*)=-S^{\uparrow}(x)+E^*E^{\uparrow}(x)+N^*N^{\uparrow}(x)
\end{equation}
where
$E^*\in \mathbb{R}$ and $N^*\in\mathbb{R}$. If we use  the notation established in the equilibrium thermodynamics, $E^*=\frac{1}{T}$
and $N^*=-\frac{\mu}{T}$, where $T$ is the equilibrium temperature and $\mu$ the equilibrium chemical potential. We want to prove that
solutions to (\ref{GEN}) approach, as $t\rightarrow\infty$ equilibrium states $\hat{x}(E^*,N^*)$ that are minima of (\ref{eq4}), i.e. that are
solutions to
\begin{equation}\label{Phimin}
\Phi^{\uparrow}_x=0
\end{equation}
We thus want to prove that solutions to (\ref{GEN}) approach the manifold
\begin{equation}\label{Meq}
\mathcal{M}^{\uparrow (eq)}=\{x\in M^{\uparrow}|\Phi^{\uparrow}_x=0\}
\end{equation}
composed of the equilibrium states $\hat{x}$.

We proceed now to recall  main steps in the proof.
If $N^{\uparrow}(x), E^{\uparrow}(x)$ is a complete set of gradient Casimirs then
the  upper reducing thermodynamic potential (\ref{eq4}) is the same as the gradient thermodynamic potential (\ref{diss2}).
Since  (due to (\ref{degen})) the Hamiltonian time evolution implies $\dot{\Phi}^{\uparrow}=0$, the upper reducing thermodynamic
potential (\ref{eq4}) plays the role of the Lyapunov function for the approach to dissipation  equilibrium
states $\mathcal{M}^{\uparrow (deq)} \equiv \mathcal{M}^{\uparrow (eq)}$ that are the same as the equilibrium states. In this case
the inequality in the third equation is sharp, the thermodynamic potential plays the role of the Lyapunov function (provided $\Phi^{\uparrow}$
is a convex function of $x$ ) and indeed, the equilibrium manifold (\ref{Meq}) is approached as $t\rightarrow\infty$.

If however the gradient part of (\ref{GEN}) has a larger set of gradient Casimirs than $N^{\uparrow}(x), E^{\uparrow}(x)$ (i.e.
if the dissipation is weaker)  then the gradient part of (\ref{GEN}) drives solutions
to
\begin{equation}\label{Meqq}
\mathcal{M}^{\uparrow (deq)}=\{x\in M^{\uparrow}|[\Xi^{\uparrow}_{x^*}]_{x^*=S^{\uparrow}_x(x)}=0\}
\end{equation}
For example (see more in Section \ref{Boltzmann}), in the
Boltzmann kinetic theory the set  $\mathcal{M}^{\uparrow (deq)}$ is composed of local Maxwell  distribution functions
and $\mathcal{M}^{\uparrow (eq)}$ of total Maxwell distribution functions. The inequality $\dot{\Phi}^{\uparrow}\leq 0$ does not suffice to
prove the approach to the manifold of equilibrium states $\mathcal{M}^{\uparrow (eq)}$. What is needed in addition is to prove that in the course
of the time evolution solutions to (\ref{GEN}) never touch $\mathcal{M}^{\uparrow (deq)}$. Only at the final destination the solution
to (\ref{GEN}) settles on both $\mathcal{M}^{\uparrow (deq)}$  and $\mathcal{M}^{\uparrow (eq)}$. This phenomenon was started to be investigated by
Grad \cite{Gradvil}. A  complete and rigorous mathematical proof  for the Boltzmann equation has earned Cedric Villani Fields Medal \cite{Villani}.
We shall refer to the  enhancement of dissipation arising in the combined gradient and Hamiltonian dynamics
\textit{Grad-Villani dissipation enhancement}.
It is very
likely an important mechanism in the onset of dissipation. Only a seed (a nucleus)  of dissipation can trigger  the   passage from  an upper
level to a lower level expressed mathematically in the reducing time evolution.

Finally, we sum up the input and the output of the reducing time evolution to the equilibrium level. The input consists of the reducing  thermodynamic
relation
\begin{eqnarray}\label{eq11}
N&=&N^{\uparrow}(x)\nonumber \\
E&=&E^{\uparrow}(x)\nonumber \\
S&=&S^{\uparrow}(x)
\end{eqnarray}
which, if inserted into (\ref{GEN}), implies that the manifold $\mathcal{M}^{\uparrow (eq)}$ given in (\ref{Meq}) is approached as
$t\rightarrow\infty$ and no time evolution takes place on the equilibrium level, i.e.
\begin{equation}\label{fixp}
\left[L^{\uparrow}E^{\uparrow}_x+\left[\Xi^{\uparrow}_{x^*}\right]_{x^*=S^{\uparrow}_x}\right]_{\mathcal{M}^{\uparrow (eq)}}\equiv 0
\end{equation}
The two potentials
$(E^{\uparrow}(x), N^{\uparrow}(x))$ have been introduced in (\ref{upeq}) and $S^{\uparrow}(x)$ in (\ref{gr1}). All three arise either from a
detailed  experimental investigation of the preparation process for the equilibrium (by trying to express it mathematically) or from a pattern
recognition process in the microscopic phase portrait (see the beginning of Section \ref{grevol}).

The equilibrium thermodynamic relation
\begin{eqnarray}\label{lowthr}
N&=&N\nonumber \\
E&=&E\nonumber \\
S&=& S(E,N)
\end{eqnarray}
is the output of the reduction. It is
obtained from (\ref{eq11})  by following the time evolution governed by (\ref{GEN}) to
its conclusion (see more in Section \ref{mp}).

\subsection{MaxEnt Passage }\label{mp}

Investigations of the process of preparation for the equilibrium level (i.e. investigations of solutions to the upper reducing
time evolution equation (\ref{GEN})) in Section \ref{GENsection} led us to  the reducing  thermodynamic relation (\ref{eq11}). We have seen that
solutions to (\ref{GEN}) approach, as $t\rightarrow\infty$,   equilibrium states $\hat{x}(E^*,N^*)\in \mathcal{M}^{\uparrow (eq)}$ that are minima
of the upper thermodynamic potential (\ref{eq4}) ( i.e. $\hat{x}(E^*,N^*)$  are solutions to (\ref{Phimin})).

We now take this result of  investigations of the process of preparation for the equilibrium level as our starting point and make the passage to
the equilibrium level without an explicit reference to the preparation process itself. We thus begin with the reducing thermodynamic
 relation (\ref{eq11})
and with the MaxEnt principle. Our objective is to pass to the equilibrium level and arrive at the equilibrium thermodynamic relation (\ref{lowthr})
implied by (\ref{eq11}). The passage (\ref{eq11}) $\rightarrow$ (\ref{lowthr}) is a mapping that, as we shall see below, is a reducing Legendre
transformation. The same mapping is made in Section \ref{GENsection} but by following the time evolution governed by (\ref{GEN}). The maximization
of the entropy $S^{\uparrow}(x)$ subjected to constraints $E^{\uparrow}(x), N^{\uparrow}(x)$, postulated in this section (MaxEnt principle)
is in Section \ref{GENsection}  a consequence of the reducing time evolution governed by (\ref{GEN}). Also the reducing thermodynamic relation
(\ref{eq11}) has arisen  in Section \ref{GENsection} from an analysis of the process of preparation for the equilibrium level. In this
section where we do not consider the preparation process we have to either postulate it or obtain it by using  arguments based on various
interpretations of the entropy (e.g. its relation to the measure of information) that were developed mainly in the context of the Gibbs
equilibrium statistical mechanics (i.e. in investigations made on the microscopic level) or in the stochastic approach to thermodynamics.

The equilibrium  thermodynamic relation $S^*(E^*,N^*)$ implied by (\ref{eq11}) is
\begin{equation}\label{eq10}
S^*(E^*,N^*)=\Phi^{\uparrow}(\hat{x}(E^*,N^*); E^*,N^*)
\end{equation}
where   $\hat{x}(E^*,N^*)$ is an equilibrium state (i.e. a solution to (\ref{Phimin})).  The equilibrium  thermodynamic relation $S=S(E,N)$ is
then obtained by the Legendre transformation. This means that
\begin{equation}\label{eq20}
S(E,N)=\Phi^*(\widetilde{E^*}(E,N),\widetilde{N^*}(E,N);E,N)
\end{equation}
where $\Phi^*(E^*,N^*;E,N) =-S^*(E^*,N^*)+E^*E+N^*N$ and $(\widetilde{E^*}(E,N),\widetilde{N^*}(E,N)) $ is a solution to
$\Phi^*_{E^*}=0, \Phi^*_{N^*}=0$.  Summing up, the equilibrium  thermodynamic relation $S=S(E,N)$  is obtained from the reducing
thermodynamic relation (\ref{eq11}) by two mappings
\begin{equation}\label{eq111}
(S^{\uparrow}(x),E^{\uparrow}(x),N^{\uparrow}(x))\rightarrow (S^*(E^*,N^*),E^*,N^*)\rightarrow (S(E,N),E,N)
\end{equation}
where the first mapping is the reducing Legendre transformation (\ref{eq10}) and the second mapping is the Legendre transformation (\ref{eq20}).
We call (\ref{eq111})  Maximum Entropy principle (MaxEnt principle).

We now comment about the physical interpretation of the quantities $E^*$ and $N^*$.  They appear  on both the upper level in the
thermodynamic potential $\Phi^{\uparrow}$ (see (\ref{eq4})) and on the equilibrium level in the equilibrium thermodynamic
potential $\Phi(E,N;E^*,N^*)=-S(E,N)+E^*E+N^*N$.

On the equilibrium level the quantities  $E^*$ and $N^*$ are the conjugate variables to $E$ and $N$ respectively since the following relations
hold: $E^*=S_E(E,N)$ and $N^*=S_N(E,N)$. They play a very important role in the equilibrium thermodynamics since they can be easily measured.
The measurement of $E^*$ (and consequently of the temperature $T$ since $E^*=\frac{1}{T}$) is made possible by the ubiquity in the nature of
membranes which either pass freely or block completely the internal energy $E$.  If a macroscopic system is put into the contact in which the internal
energy freely passes with a thermometer (which is another macroscopic system) and both the system and the thermometer are surrounded by the
membrane that blocks the passage of the internal energy then, due to the maximization  of the entropy in the established equilibrium states, the
temperature of the system becomes the same as the temperature of the thermometer. The temperature of the thermometer is then made visible
through a known relation between the temperature and  another state variable of the thermometer (e.g. volume or pressure) that can be directly
observed. The  existence of membranes that freely pass or block the passage of the mass then similarly makes possible  to measure $N^*$.
Moreover, since $E^*=S_E>0$, there is a one-to-one relation between the equilibrium thermodynamics formulated in terms
of $(S(E,N),E,N)$ and $(E(S,N),S,N)$. Using the terminology of Callen \cite{Callen}, the former formulation is called an entropy representation
and the latter an energy representation of the equilibrium thermodynamics.

On the upper level the quantities
$(E^*,N^*)$  play only the role of  the Lagrange multipliers in the maximization of the reducing  entropy $S^{\uparrow}(x)$. They are not anymore
conjugate variables and they cannot be simply measured on the upper level. This is the well known problem with the definition and measurements of
the temperature on mesoscopic levels (including the levels used in direct numerical simulation).

The observations that we just made about $(E^*,N^*)$ are also related to the relation between the entropy representation
(in which $(E,N)$ are independent
state variables) and the energy representation  (in which $(S,N)$ are independent state variables) in the classical equilibrium
thermodynamics (see \cite{Callen}). Due to the positivity of the absolute temperature $T=(S_E(E,N))^{-1}$, these two representations
are interchangeable in the classical equilibrium thermodynamics. The equilibrium fundamental thermodynamics relation can be given either
in the form (\ref{lowthr}) or in the form ($N=N,E=E(S,N),S=S$). This exchangeability of entropy and energy representations extends to fluid
mechanics (with fields of mass, energy and momentum playing the role of state variables) only under the local equilibrium assumption accorging
to which the entropy field (i.e. the local entropy) is the same function of the mass and the energy fields as in equilibrium
and thus the field of the temperature (i.e. the local temperature) is positive.
In the context of a general mesoscopic level with state variables $x$,  the upper reducing
thermodynamic relation has only one form (\ref{eq11}). There are  no energy and  entropy representations.

It is also interesting to note the difference in the inclusion  of constraints in the maximization of the reducing  entropy $S^{\uparrow}(x)$
made in the MaxEnt principle and in the maximization of the same entropy  made by following the reducing time evolution. While the former is
made simply by the method of Lagrange multipliers the latter, as we have seen in Section \ref{GENsection}, is made by requiring degeneracies of
the geometrical structures involved in the vector fields and by proving the Grad-Villani dissipation enhancement.

Still continuing with the comparison of the MaxEnt reduction (in this section) and the reduction  made in the reducing time evolution
(in Section \ref{GENsection}, we look more closely into the role of Legendre transformations. We have already noted
that the MaxEnt reduction (\ref{eq111}) is a
sequence of two Legendre transformations. The first one is a reducing Legendre transformation and the second is a regular Legendre transformation.
A natural question is as to whether the reducing time evolution is  in fact also a sequence (an infinite sequence)  of (infinitesimal)
Legendre transformations.
We answer this question in the next section.

\subsection{Contact Geometry}\label{CS}

Having realized that the fundamental group of thermodynamics is the group of Legendre transformations, we ask the question of what is the
mathematical environment in which the Legendre transformations appear as natural transformations.
The geometrical structure that is preserved in the Legendre transformations is the contact structure \cite{Arnoldmech}, \cite{construc}.
We can thus suggest that the
contact geometry provides a natural mathematical environment for thermodynamics. For the classical equilibrium thermodynamics this suggestion
was  made in \cite{Hermann}, \cite{Mrugala} and for the multiscale thermodynamics in \cite{Grmelacont}.
In this section we only discuss the physical aspects
of the contact-geometry formulation of thermodynamics. Its   mathematical background can be found in \cite{Arnoldmech}, \cite{construc}.

The time evolution governed by
(\ref{GEN}) will be a sequence of Legendre transformations if   (\ref{GEN}) is lifted into a larger space that is equipped with a contact structure
and the lifted equation  (\ref{GEN})  will  generate the time evolution that preserves the contact structure. From the side of physics,
the motivation (and guidance) for this type of reformulation of (\ref{GEN}) comes from the following considerations. In the classical (both
equilibrium and nonequilibrium)
thermodynamics the conjugate state variables  like the temperature and the pressure
 play the role of the same (if not larger) importance as the
energy and volume. We can therefore suggest
to  adopt the conjugate state variables $x^*$ as independent state variables. We introduce a large
space $\mathbb{M}^{\uparrow}$  with  coordinates
$(x,x^*,z)$ , $x\in M^{\uparrow};
x^*\in M^{\uparrow *}; z\in \mathbb{R}$.
The fact that $x$ and $x^*$
are related in thermodynamics by $x^*=S^{\uparrow}_x(x)$ suggests that its submanifold
\begin{equation}\label{Legman}
\mathfrak{M}^{\uparrow}=\{(x,x^*,z)\in \mathbb{M}^{\uparrow}|x^*=S^{\uparrow}_x(x); z=S^{\uparrow}(x)\}
\end{equation}
is both physically and  mathematically significant. From the physical point of view, the
thermodynamics takes place on
$\mathfrak{M}^{\uparrow}$. The  mathematical significance of the submanifold $\mathfrak{M}^{\uparrow}$ stems from the fact that
$\mathbb{M}^{\uparrow}$ is equipped with    the contact structure defined by the 1-form
\begin{equation}\label{1form}
dz-x^*dx
\end{equation}
and $\mathfrak{M}^{\uparrow}$ is its
a Legendre submanifold. We recall that
the Legendre submanifold
 is defined as a manifold on which the contact 1-form equals zero.
We note that $[dz-x^*dx]_{\mathfrak{M}^{\uparrow}}=0$.

In order to include the MaxEnt reduction (\ref{eq11}) into  the contact geometry formulation  we  have to still enlarge the space
$\mathbb{M}^{\uparrow}$. We enlarge it into the space
$\widehat{\mathbb{M}^{\uparrow}}$ with
coordinates $(x,x^*,E^*,N^*,E,N,z)$ and equip it with the 1-form $dz-x^*dx-EdE^*-NdN^*$. The Legendre manifold (\ref{Legman}) turns
in $\widehat{\mathbb{M}^{\uparrow}}$   into another  Legendre manifold
\begin{equation}\label{LLegman}
\widehat{\mathfrak{M}^{\uparrow}}=\{(x,x^*,E^*,N^*,E,N,z)\in \widehat{\mathbb{M}^{\uparrow}}|x^*=\Phi^{\uparrow}_x; E=\Phi^{\uparrow}_{E^*};
N=\Phi^{\uparrow}_{N^*}; z=\Phi^{\uparrow}\}
\end{equation}
The MaxEnt reduction takes place on $\widehat{\mathfrak{M}^{\uparrow}}$ and $\widehat{\mathfrak{M}^{\uparrow}}$ is again the Legendre manifold.

What remains is to lift (\ref{GEN}) to $\widehat{\mathbb{M}^{\uparrow}}$ in such a way that: (i)  the 1-form $dz-x^*dx-EdE^*-NdN^*$ is preserved in the time evolution generated by the lifted (\ref{GEN}), and (ii) the Legendre manifold
$\widehat{\mathfrak{M}^{\uparrow}}$ is invariant in the time evolution generated by the lifted (\ref{GEN}) and the lifted equation
(\ref{GEN}) restricted to $\widehat{\mathfrak{M}^{\uparrow}}$ is exactly the equation (\ref{GEN}). The time evolution
in $\widehat{\mathbb{M}^{\uparrow}}$ satisfying these properties will be called contact reducing time evolution.

As for the first point, the canonical form of the time evolution equations preserving a given (maximally non-integrable) 1-form is well
known \cite{Arnoldmech}, \cite{construc}. The form resembles the form of the canonical Hamilton equations. In particular, the vector field is a gradient of a
potential (called a reducing contact Hamiltonian $\mathbb{E}^{\uparrow}(x,x^*,E^*,N^*,E,N,z)$) transformed into a vector by the contact geometrical
structure (similarly
as the Hamiltonian vector field
(\ref{Ham1}) is the gradient $E^{\uparrow}_x$ of the Hamiltonian $E^{\uparrow}(x)$ transformed into a vector by the symplectic
structure, i.e. by  the bivector $L^{\uparrow}$).

Regarding the second point, the contact Hamiltonian $\mathbb{E}^{\uparrow}(x,x^*,E^*,N^*,E,N,z)$), identified in \cite{Grmelacont}, \cite{Gcon}
 is essentially
the rate reducing thermodynamic potential (\ref{ratePsi}) with $\Sigma^{\uparrow}
=\Xi^{\uparrow}(x,x^*)-\left[\Xi^{\uparrow}(x,x^*)\right]_{x^*=\Phi^{\uparrow}_x}$, $W^*=E^*$, and $W^{\uparrow}=<x^*,L^{\uparrow}E^{\uparrow}_x>$.

Summing up: (i) the contact structure of the space $\widehat{\mathbb{M}^{\uparrow}}$ remains unchanged during the contact reducing  time evolution,
(ii) the  contact reducing time evolution takes place on the Legendre manifold $\widehat{\mathfrak{M}^{\uparrow}}$ given in (\ref{LLegman}),
(iii) the geometrical structures appearing in (\ref{GEN}), i.e. the symplectic structure
expressed in the bivector $L^{\uparrow}$
and the generalized gradient structure expressed in the reducing dissipation potential $\Xi^{\uparrow}$, make their appearance in the contact
reducing time evolution in the reducing contact Hamiltonian  $\mathbb{E}^{\uparrow}(x,x^*,E^*,N^*,E,N,z)$).

The contact formulation is thus very satisfactory both from the physical and the mathematical point of view. The physical
satisfaction comes from  seeing
the reducing thermodynamic relation (\ref{eq11}) as a relation determining the manifold (the Legendre manifold (\ref{LLegman})) on which
the time evolution takes place and seeing the symplectic and the gradient geometrical structures in the
generating potential  $\mathbb{E}^{\uparrow}(x,x^*,E^*,N^*,E,N,z)$
(we recall the they enter GENERIC (\ref{GEN}) in
the geometry used to transform gradients of potentials into forces).
The mathematical satisfaction comes mainly from the fact
that the contact geometry of the space $\widehat{\mathbb{M}^{\uparrow}}$ in which the contact time evolution takes place remains unchanged during
the time evolution. From the mathematical point of view, we are thus as comfortable  as we are with the  Hamiltonian dynamics
in the setting of the symplectic geometry  and with the
gradient dynamics in the setting of the  Riemannian geometry. The GENERIC dynamics formulated in (\ref{GEN}) involves two geometrical structures
(symplectic and Riemannian),  neither of them are preserved in the course of the time evolution.

Finally, we also recall that the variational formulation that is  well known  for both the Hamiltonian dynamics and the gradient dynamics
can be, in the setting of the contact geometry, extended  to their combination, i.e. to the  GENERIC dynamics \cite{Grmelacont}. The contact
geometry provides also a natural setting for using the thermodynamic methods in the control theory \cite{Maschke}, \cite{Hudon}.

\subsection{Passage to a Lower Level with Lower Dynamics}\label{ll}

So far,  the lower level \textit{l} with which we are   comparing the upper level $\mathcal{L}$ was the equilibrium level that  distinguishes itself
among   other levels mainly by the absence of the time evolution. No time evolution  takes place on the equilibrium level (see (\ref{fixp}). We
replace now the
equilibrium level with a general (but still lower than $\mathcal{L}$) level \textit{l} on which a time evolution (called a lower time evolution)
takes place.
What does have to be changed in the investigation of the passage $\mathcal{L} \rightarrow $ \textit{l}?

If both levels $\mathcal{L}$ and \textit{l} are well established (i.e. well tested with results of experimental observations) then there has
to be a way to prepare the macroscopic systems under investigations for the level \textit{l} and such preparation process has to be
presentable as a time evolution on the level $\mathcal{L}$.  In this respect the replacement of the equilibrium level
with the level \textit{l} that involves the time evolution does not bring any change. The question that remains to be answered
is  as to whether the preparation process is governed  again by
(\ref{GEN}). We shall assume that it is (\ref{GEN}) that governs the preparation process but (\ref{GEN}) with different potentials
and geometrical structures. We recall that Eq.(\ref{GEN}) describing the preparation process to the equilibrium level has arisen as a common
structure of this type of equations developed independently, by many researchers, in different times, and  on many different levels
(see Section \ref{realizations}). No
such pool of equations  is available for investigating the approach to mesoscopic levels \textit{l} with the time evolution.
However, the basic physics that
is behind (\ref{GEN}) remains the same. We are
looking essentially at the same preparation process except that we are interrupting  it before its completion.  The microscopic
basis of the time evolution describing the preparation process
is again  the particle Hamiltonian mechanics and the gradual disappearance of details in the preparation process that is
expected to be mathematically manifested  in the gradual decrease (or increase) of a Lyapunov like potential. As for the question of what are
the potentials and the geometrical structures appearing in (\ref{GEN}) that represents a given macroscopic systems, we leave it at this point
unanswered. We shall discuss some examples in Section \ref{realizations} and Section  \ref{hierarchy}.

There is however an important  difference in the time evolution representing the preparation for the equilibrium level and for the mesoscopic level
\textit{l} involving the time evolution.
In the reduction to the equilibrium level,  solutions to the upper reducing time evolution equations approach fixed points (see (\ref{fixp})).
This means that
the fixed points are eventually (as $t\rightarrow \infty$) reached and then never leave it. The fixed points are, of course, invariant manifolds.
In other words, the approach to fixed points is automatically  an approach to an invariant manifold. In investigations of the approach to a lower
level with the lower time evolution,  trajectories in $M^{\uparrow}$ approach $\mathcal{M}^{\uparrow (low)}\subset M^{\uparrow}$ that is in one-to-one
relation with the lower state space $M^{\downarrow}$.

The requirement of the  invariance of the manifold  $\mathcal{M}^{\uparrow (low)}$ is now highly non trivial.
It is this requirement
that makes the investigation of the reduction to a lower level with the time evolution more difficult than  the investigation of the reduction to
the equilibrium level. The result of the investigation $\mathcal{L}\rightarrow$ \textit{l}, where \textit{l} involves the lower time evolution,
is not only the lower thermodynamic relation (i.e. the equilibrium thermodynamic relation when \textit{l} is the equilibrium level) but also
the lower time evolution.
In historically the first investigation of this type \cite{Chapman-Enskog}, known as the Chapman-Enskog method,  the level $\mathcal{L}$ is the
level of kinetic theory represented by the Boltzmann kinetic equation and the level \textit{l} is the level of hydrodynamics with the five
hydrodynamic fields serving as state variables (see more in Section \ref{Boltzmann}).

As in the investigation of the reduction to the equilibrium level (see (\ref{upeq})) we  begin with
\begin{equation}\label{uppeq}
M^{\uparrow}\rightarrow M^{\downarrow (l)};\,\,x\mapsto y(x))
\end{equation}
In the  Chapman and Enskog investigation (see more in Section \ref{Boltzmann}),
$y(x)$ are the hydrodynamic fields expressed in terms of the one particle distribution function.

The reducing time evolution equation is (\ref{GEN})  with the reducing thermodynamic relation
\begin{eqnarray}\label{redloweq}
&& y=y(x)\nonumber \\
&&E^{\uparrow}(x)\nonumber \\
&&S^{\uparrow}(x)
\end{eqnarray}
Both the reducing energy $E^{\uparrow}(x)$ and the reducing entropy $S^{\uparrow}(x)$ are, in general, different from those
appearing in (\ref{eq11}). All reductions
depend on both the upper level $\mathcal{L}$ and the lower level \textit{l}. If the lower level changes all quantities appearing
in the reduction change. In particular the energy  $E^{\uparrow}(x)$ appearing in (\ref{redloweq}) is  only the energy involving the state
variables that belong to $x$ but do not belong to $y$.
In order to avoid overburdening our notation, we do not show explicitly the  dependence on the lower level \textit{l}.

The reducing thermodynamic relation (\ref{redloweq}) is again
obtained as a result of a pattern recognition process in the upper phase portrait but the focus is put on a different pattern than in the
the investigation of the passage to the equilibrium level. Instead of looking for the fixed points (\ref{Meq}),
we look for  manifolds  $\mathcal{M}^{\uparrow (low)}$ satisfying the following five properties:
\begin{eqnarray}\label{mlow}
&&\,\,\,\, (i) \mathcal{M}^{\uparrow (low)}\subset M^{\uparrow}\nonumber \\
&&\,\,\,\, (ii)\mathcal{M}^{\uparrow (low)}\,\,is\,\,in\,\,one-to-one\,\,relation\,\,with\,\,M^{\downarrow}\nonumber \\
&&\,\,\,\, (iii) \mathcal{M}^{\uparrow (low)}=\{x\in M^{\uparrow}|\Phi^{\uparrow}_x(x;y^*)=0\}\nonumber \\
&&\,\,\,\, (iv) \mathcal{M}^{\uparrow (low)}\,\,is\,\,approached\,\,as\,\,t\rightarrow\infty\nonumber \\
&& \,\,\,\,(v) \mathcal{M}^{\uparrow (low)}\,\, is\,\, maximally\,\, invariant
\end{eqnarray}
where
\begin{equation}\label{Phil}
\Phi^{\uparrow}(x;y^*)=-S^{\uparrow}(x)+E^{\downarrow *}E^{\uparrow}(x)+<y^*,y(x)>
\end{equation}
In the context of the reducing time evolution equation (\ref{GEN}) we require that $y(x)$ is both the Casimir and the gradient
Casimir, $E^{\uparrow}(x)$ is the gradient Casimir
and $S^{\uparrow}(x)$ is the Casimir. The fifth requirement was not,  of course, needed in the previous section. This new requirement
plays now a very important role in determining the potentials appearing
in (\ref{redloweq}). The precise meaning of \textit{maximally invariant} (or alternatively  "quasi-invariant") used in (\ref{mlow})
as well as the meaning of "appropriately projected"
used in (\ref{fixppp}) below remains still a
part of the pattern-recognition analysis of the upper time evolution that has to be investigated  \cite{GorbanKarlin},
\cite{Grcheming}, \cite{red1}, \cite{red2}, \cite{red3}.

The output of the reduction $\mathcal{L}\rightarrow  l$  is the lower thermodynamics relation
\begin{equation}\label{lowthr1}
S^{\downarrow}=S^{\downarrow}(E^{\downarrow}, y)
\end{equation}
(obtained from (\ref{redloweq}) in the same way as (\ref{lowthr}) is obtained from (\ref{eq11}) -  see more in Section \ref{mp}) and
the vector field (compare with (\ref{fixp}))
\begin{equation}\label{fixppp}
\left[L^{\uparrow}E^{\uparrow}_x+\left[\Xi^{\uparrow}_{x^*}\right]_{x^*=S^{\uparrow}_x}\right]_{\mathcal{M}^{\uparrow (low)}}
\end{equation}
that,  if appropriately projected on the tangent space $T\mathcal{M}^{\uparrow (low)}$ of the manifold $\mathcal{M}^{\uparrow (low)}$ and pushed
forward on $M^{\downarrow (l)}$ by the mapping (\ref{uppeq}),   becomes  the vector field generating   the time evolution on the lower level \textit{l}.
 In this paper we limit ourselves only to
recalling the main idea behind the  Chapman and Enskog analysis (see more in Section \ref{Boltzmann} and in Refs. \cite{Chapman-Enskog},
\cite{GorbanKarlin},  \cite{Grcheming},  \cite{red1}, \cite{red2}, \cite{red3})..

Before leaving this section we make two remarks.

\textit{Remark 1}

The original Chapman-Enskog investigation of the reduction of kinetic theory to hydrodynamics, as well as its
continuation in \cite{GorbanKarlin}, concentrate only on the derivation of the
lower time evolution generated by the vector field (\ref{fixppp}). The larger context of multiscale reductions has led us to the derivation
of an additional result, namely to the reduced thermodynamics relation (\ref{lowthr1}) that is associated with
the lower time evolution. In the reduction $\mathcal{L}\rightarrow$ \textit{equilibrium level} we have  obtained the equilibrium thermodynamic
relation as  the reduced thermodynamic relation and the reduced vector field is no vector field.
In the reduction $\mathcal{L}\rightarrow  l$  that involves time evolution
 we obtain lower thermodynamic relation (\ref{lowthr1}) and lower time evolution generated
by (\ref{fixppp}).
The reduced thermodynamic relation (\ref{lowthr1}) represents the thermodynamics on the equilibrium level that is inherited form the
reduction $\mathcal{L}\rightarrow$ \textit{equilibrium level}.
The reduced thermodynamic relation (\ref{lowthr1}) represents the thermodynamics on the level \textit{l} that is inherited from the
reduction $\mathcal{L}\rightarrow  l$. As we saw in  Section \ref{CS}, the reduced thermodynamic
relation (\ref{lowthr}) provides the lower state space $M^{\downarrow (i)}$ with geometry.

\textit{Remark 2}

Externally or internally driven macroscopic systems are prevented from reaching the equilibrium level. The equilibrium thermodynamics does
not exist for such systems. However, the behavior of  externally or internally driven macroscopic systems
can often be  described
on a mesoscopic level \textit{l}. For example the experimentally observed behavior of the Rayleigh-B\'{e}nard system (a horizontal layer of a fluid
heated from below) can be described on the level of hydrodynamics (with Boussinesq equations governing the lower time evolution). In other words,
the level of hydrodynamics is well established for the  Rayleigh-B\'{e}nard system. This then means that any other level $\mathcal{L}$
 that involves  more details and that allows to express
the physics of  the Rayleigh-B\'{e}nard system (for example the microscopic level) has to be reducible to the level of hydrodynamics. The resulting
lower thermodynamic relation (\ref{lowthr1}) implied by the reduction  provides thus thermodynamics replacing the equilibrium thermodynamics
that does not exist.  Summing up, if there exists a well established mesoscopic level for an externally or internally driven macroscopic system
(however far from equilibrium and however strong are the external and the internal driving forces) then
there also exists thermodynamics (expressed in the thermodynamic relation (\ref{lowthr1})) for such system.

\subsection{Transitivity of Reductions}\label{trans}

A single  reduction $\mathcal{L}\rightarrow l$ introduces two structures: \textit{reducing structure} on the  level $\mathcal{L}$
and \textit{reduced structure } on the level \textit{l}. The reducing structure consists of the reducing thermodynamic relation and the
reducing time evolution equation. The reduced structure consists of the reduced thermodynamic relation and the reduced time evolution
equation. Moreover, since
for a given lower level \textit{l}  there
are, in general, many upper levels $\mathcal{L}$ from which it can be reduced, every level has  not one but many reducing and reduced structures.
In addition, by replacing the reduction with rate reduction, the number  of the structures is multiplied by two.

Not all such structures are however independent. We shall now explore  some of the relations among them. First, we turn to systems with
no external and internal forces that would prevent
approach to equilibrium.
and to the chain
\begin{equation}\label{chain1}
\mathfrak{L}\longrightarrow \mathcal{L}\longrightarrow equilibrium
\end{equation}
We expect that
the reductions are transitive in the sense that  the reduced equilibrium structure arising as a result of the gradual reduction
$\mathfrak{L} \longrightarrow \mathcal{L} \rightarrow equilibrium$ is the same as the reduced equilibrium structures obtained
from the direct reductions
$\mathfrak{L} \longrightarrow equilibrium$ and  $ \mathcal{L} \longrightarrow equilibrium$. This transitivity then implies the following relation
between the reduced and the reducing entropies on the level $\mathcal{L}$:
\begin{equation}\label{333}
H^{\downarrow }(y)=S^{\uparrow}(y)
\end{equation}
where $H$ denotes  the entropy associated with the reduction $\mathfrak{L}$ $\longrightarrow \mathcal{L}$ and $S$ is the entropy
associated with the reduction $\mathcal{L} \longrightarrow$ \textit{equilibrium}.

Gradual reductions (\ref{chain1}) are  more difficult to investigate than  direct reductions. Nevertheless,
we can, at least partially, illustrate the relation (\ref{333}) with two examples. In both examples
the upper level  $\mathfrak{L}$  is the level of kinetic theory.
The intermediate  level  $\mathcal{L}$ is in the first example the level of the classical
fluid mechanics with the fields of mass, momentum, and internal energy
as state variables. In the second example the intermediate level $\mathcal{L}$ is
the level of  the  extended fluid mechanics  with $n$ fields, that are velocity moments of
the one particle distribution function   (see more in Section \ref{hierarchy}).

In both examples only the nondissipative part of the  time evolution on the level $\mathcal{L}$
is considered. The passage $\mathfrak{L} \longrightarrow \mathcal{L}$ is,  in
both examples,  the MaxEnt reduction (see
 Section \ref{mp}) which does not explicitly involve the reducing time evolution.
The reduction
$\mathcal{L} \longrightarrow$ \textit{equilibrium} is not made, in both examples,  in the way described
in Section \ref{GENsection} but in the way developed in the classical nonequilibrium
thermodynamics  (i.e. as an appearance of a companion local conservation law implied by a system of local
conservation laws - see  Section \ref{lcl}).

The first example is in fact a well known result of the classical nonequilibrium thermodynamics. We begin with the reducing thermodynamic relation
(\ref{redloweq}) on the level $\mathfrak{L}$ in which
$x= f(\rr,\vv);\,\,\, y(x)= (\rho(\rr),\uu(\rr),e(\rr)) = (\int d\vv f, \int d\vv \vv f; \int d\vv \frac{\vv^2}{2});\,\,\,
N^{\uparrow}(x)=\int d\rr\int d\vv f;\,\,\, E^{\uparrow}(x)=\int d\rr\int d\vv \frac{\vv^2}{2};\\
H^{\uparrow}(x)=-\int d\rr\int d\vv f\ln f$.
The fields $(\rho(\rr),\uu(\rr),e(\rr))$ are hydrodynamics fields, $\rho$ is the mass, $\uu$ momentum and $e$ internal energy.
The reducing thermodynamic potential (\ref{eq4}) reaches its minimum at the local Maxwell distribution and finally the  reduced
entropy $S^{\downarrow}(y)$ on the level $\mathcal{L}$ is the local equilibrium entropy  given in (\ref{lowthr1}). Neither the reducing
nor the reduced time evolution is included in the investigation of the reduction  $\mathcal{L}\rightarrow l$.

Next, we look at  the time-evolution passage  $\mathcal{L} \longrightarrow equilibrium$. The nondissipative part of the reducing time evolution
is the Euler hydrodynamics.
Together with the reducing thermodynamic relation (\ref{eq11}) in which  $N=\int d\rr \rho(\rr);\,\,E=\int d\rr e(\rr)$, and
$S=\int d\rr s(\rho,\uu,e;\rr)$, where
$s(\rho,\uu,e;\rr)$ is the local equilibrium entropy field,  the Euler hydrodynamic equations
implies (\ref{gen1}) with the equality in the third equation. The entropy conservation $\dot{S}^{\downarrow}=0$ arises in fact
 as a local conservation law $\frac{\partial s}{\partial t}=-\frac{\partial \left(s u_i/\rho\right)}{\partial r_i}$ (see more in Section \ref{fmt}).
This is a well known result of the classical nonequilibrium thermodynamics.
The relation (\ref{333}) is thus in the context of the first illustration   proven.

In  the second example \cite{Dreyer}, we put the first example  into the larger context of
Grad's hierarchy that is a particular reformulation of Boltzmann's kinetic equation in which the one particle
distribution function is presented in the form of an infinite set of equations governing the time evolution of
its velocity moments (see more in Section \ref{hierarchy}).
We realize that the hydrodynamic fields are the first five
moments. In the context of Grad's hierarchy, the  first illustration is in fact a splitting the infinite Grad hierarchy
into two parts: the lower part  is a closed system
of  equations governing the first five moments (the governing equations of Euler's hydrodynamics) and the upper part are the remaining
equations in the infinite hierarchy. The analysis of solutions to the upper part of the hierarchy is replaced (as it was done also in the first
illustration) by the
MaxEnt passage (with the Boltzmann entropy)  from the one particle distribution to its five moments.
The second illustration of (\ref{333}), worked out  in \cite{Dreyer},  is thus the same as the first one
 but with a general number $n$ of Grad's moments serving as hydrodynamic fields.
 Dreyer proves in \cite{Dreyer}   that all the results that we have recalled above in the first example for $n=5$ hold also for $n>5$.

Transitivity of rate reductions in the chain (\ref{chain1}) is discussed in Section \ref{rt}.

\subsection{Criticality}\label{crit}

The strength of the  autonomy of a level is  measured by the strength  of fluctuations. The larger are the fluctuations the less autonomous is
the level. Large fluctuations indicate that the details that were ignored, in both experimental observations and in the mathematical formulation,
cannot be  ignored anymore.
From the mathematical point of view, the loss of autonomy is manifested by  the  loss of convexity of
thermodynamic potentials.  The regions in which this is happening are called critical regions.

Investigations of critical phenomena bring extra difficulties but also extra simplifications. The first simplification is the mathematical
universality of reducing thermodynamic potentials. As shown in the catastrophe theory \cite{Arnoldcat}, real
 valued smooth functions have in the vicinity of their
degenerate critical points only a few nonequivalent forms (Landau polynomials).
 The second simplification
is the inseparability of levels in the critical region allowing to define the critical region alternatively in terms of reductions.

The first simplification was noted  by Landau \cite{Landau}. Viewing his theory through the eyes of multiscale thermodynamics, we formulate it
in the following   three steps. (i) The equilibrium level is extended to an upper level with an order parameter, $\xi$,  playing the role of an
extra state variable,  (ii) The upper reducing thermodynamic potential potential $\Phi(\xi;E^*,N^*)$ has a universal form of  Landau polynomials
in the extra state variable $\xi$
\cite{Arnoldcat}.
(iii) MaxEnt reduction of the extension formulated on the upper level to the equilibrium level implies a universal critical behavior
 at the equilibrium level.

 This multiscale viewpoint of the
 Landau theory is formulated and illustrated on the van der Waals theory in \cite{GrPhysA} \cite{LanGrPK}. The mathematical results of the
 universality of thermodynamic potentials in critical regions, that have arisen in the catastrophe theory \cite{Arnoldcat},  may become less
 surprising if we allegorically compare them with the very familiar observation that our two friends, Bill and Bob,
 are very different in many respects but their behavior  in critical situations is very similar. The criticality overrides the diversity.

The realization of the existence  of the second simplification has  originally arisen in  the comparison of   predictions of
the Landau theory with results
of experimental observations. The agreement is found to be only qualitative. In order to explain it, the attention was turned to the
inseparability of levels at the critical point. It has been realized that the critical points themselves  can be defined as  fixed points of a
group of transformations (called a
renromalization group ) representing a pattern recognition process, The fixed point, i.e. the critical point, is the point where no pattern can
be recognized. This type of idea has  originally  been formulated in the context of the Gibbs equilibrium statistical mechanics in \cite{Wilson}.
In this formulation the microscopic Hamiltonians approach in the pattern recognition process (consisting usually of a spatial coarse graining)
to fixed points.
In the context of the multiscale thermodynamics, the renormalization-group  approach to critical phenomena has been
formulated in \cite{GrPhysA} \cite{LanGrPK}. In this formulation the coefficients of the Landau polynomials approach the fixed points. The
pattern recognition process is not the spatial coarse graining but an extension of the original  one component systems to
a two component system  followed by MaxEnt reduction back to the original one component system. The two components are complete identical, they
are distinguished only by a  feature that does not influence at all the physical properties determining dynamics (e.g. by a colour).

\section{Rate Thermodynamics}\label{rt}

As we have already emphasized several times, the reduction $\mathcal{L} \rightarrow$ \textit{l} is either a mathematical formulation of the
experimental investigation of the preparation process for the level \textit{l} or, in the case when the time evolution taking place on the
level $\mathcal{L}$ is known,  a pattern recognition process in the upper phase portrait. The recognized pattern is then the reduced phase
portrait. In the search for the pattern we have so far concentrated on the phase portrait in the state space $M^{\uparrow}$. Alternatively, we can
look what is happening in the course of the preparation process  in the space $\mathfrak{X}(M^{\uparrow})$ of the vector fields on $M^{\uparrow}$.
Such change in  the focus of our attention is expected to help in  recognizing  overall features since the lift  to higher order
tangent spaces is in fact a way to see  larger pieces of  trajectories.  Moreover, the recognized lower time evolution will appear
in $\mathfrak{X}(M^{\uparrow})$ as a fixed point (as the lower vector field) and not as a quasi-invariant submanifold of the
state space $M^{\uparrow}$. It is easier to recognize fixed points than quasi-invariant submanifolds.
For the reason that will appear later in the discussion of relations between rate reductions and reductions
(see also (\ref{disspot}))  we shall observe the upper time
evolution in the space $\mathfrak{X}^*(M^{\uparrow})$ of co-vector fields rather than in the space $\mathfrak{X}(M^{\uparrow})$
of vector fields. The elements
of $\mathfrak{X}^*(M^{\uparrow})$  (denoted by the symbol $X$ , i.e. $X\in \mathfrak{X}^*(M^{\uparrow})$)  are physically
interpreted as thermodynamic forces.

The change from $M^{\uparrow}$ to  $\mathfrak{X}^*(M^{\uparrow})$  is reflected in our terminology by adding the prefix "rate". The reducing time
evolution in  $\mathfrak{X}(M^{\uparrow})$ is thus reducing rate-time evolution and  the thermodynamic relation is rate thermodynamic relation.
The elements of  $\mathfrak{X}^*(M^{\uparrow})$  are denoted $X$ (i.e. $X\in  \mathfrak{X}^*(M^{\uparrow})$ ) and $Y\in  \mathfrak{X}^*(M^{\downarrow})$.
The reducing rate-entropy  is denoted $\Sigma^{\uparrow}(X)$, the reducing rate-energy   $W^{\uparrow}(X)$, and the
reducing rate-thermodynamic potential  $\Psi^{\uparrow}(X;I)$.
Similarly, $\Sigma^{\downarrow}(Y)$ is the reduced rate-entropy and $W^{\downarrow}(Y)$ the reduced rate-energy.
The Maximum Entropy principle (MaxEnt principle) becomes Maximum
Rate Entropy principle (MaxRent principle).  We try to use the notation established in nonequilibrium thermodynamics (see also
Section \ref{fmm}). We depart therefore from using $X^*$ to denote the conjugate of $X$. Instead, we denote the  conjugates of $X$, having the  physical
interpretation of  thermodynamic fluxes, by  the symbol $J$ as it is customary  in the classical nonequilibrium thermodynamics (see also
Section {fmm}).
Similarly, on the lower level \textit{l}, the thermodynamic forces are denoted by the symbol $Y$ and its conjugates,
having the physical interpretation of lower level thermodynamic fluxes, by the symbol $I$.

From the physical point of view, the preparation process for the
level \textit{l} is the same as in Section \ref{ll}. We just observe it differently.
The  mathematical formulation of the MaxRent passage $\mathcal{L} \rightarrow l$ begins with
the reducing rate thermodynamic relation
\begin{eqnarray}\label{rateenergy}
&&Y= Y(X)\nonumber \\
&&W^{\uparrow}(X)\nonumber \\
&&\Sigma^{\uparrow}(X)
\end{eqnarray}
where $W^{\uparrow}(X)$ is a rate of energy.
The corresponding to it
rate thermodynamic potential reads
\begin{equation}\label{ratePsi}
\Psi^{\uparrow}(X;W^{\downarrow *},I)=-\Sigma^{\uparrow}(X)+W^{\uparrow}(X)W^{\downarrow *}+<Y(X),I>
\end{equation}
where $(W^{\downarrow *}, I)$ are Lagrange multipliers.

Next, we pass by the MaxRent reduction from $\Sigma^{\uparrow}(X)$ to
$\Sigma^{\downarrow *}(W^{\downarrow *},I)$
and finally (by the ordinary Legendre transformation) to $\Sigma^{\downarrow }(W^{\downarrow },Y)$. The Legendre transformations involved
in the MaxRent reduction are the same as the Legendre transformations made in the MaxEnt reduction in Section \ref{mp}.

If we compare the rate thermodynamic potential (\ref{ratePsi})
with the thermodynamic potential (\ref{eq4}), we note that the coefficient  $W^{\downarrow *}$ is in rate reductions analogous to
the coefficient $E^*=\frac{1}{T}$ introduced
in reductions. We therefore interpret physically $W^{\downarrow *}$ as an inverse \textit{rate temperature} $\mathcal{T}$, i.e.
$W^{\downarrow *}=\frac{1}{\mathcal{T}}$. In terms of the lower entropy $\Sigma^{\downarrow }(W^{\downarrow },Y)$, the rate temperature $\mathcal{T}$
becomes
\begin{equation}\label{ratetemp}
\Sigma^{\downarrow }_{W^{\downarrow}}(W^{\downarrow },Y)=\frac{1}{\mathcal{T}}
\end{equation}
The rate temperature $\mathcal{T}$ can be measured with a rate thermometer similarly as the temperature $T$ is measured with a
standard thermometer. The
difference between the standard   and the rate thermometers is in the walls separating the thermometers from the system whose
temperature is measured. In the standard  thermometers it is a wall that freely passes or stops passing the internal energy. Such walls
are ubiquitous in the nature. In the rate thermometers the walls have to freely pass or stop passing the rate of the internal energy.
Such walls are certainly not ubiquitous in the nature.  The rate temperature remains thus still only a theoretical concept.

As for the rate time-evolution passage $\mathcal{L} \rightarrow$ \textit{l},
we have already noted
that even if the lower level
\textit{l} involves the lower time evolution in the space of vector fields $\mathfrak{X}^*(M^{\uparrow})$  still approaches to a
fixed point (to the lower vector field) and not to a quasi-invariant manifold.
We shall make some additional observations
addressing this aspect of the reducing rate time-evolution  in Section \ref{hierarchy}. Regarding other aspects of the time evolution
governing the passage
$\mathfrak{X}^*(M^{\uparrow})\rightarrow \mathfrak{X}^*(M^{\downarrow})$, we conjecture that  it
possesses the GENERIC structure discussed in Section \ref{GENsection}. Contrary to the passage $M^{\uparrow}\rightarrow M^{(eq)}$ for which we have
many specific examples (that have been developed independently and on many different levels) that all possess the GENERIC structure, the
argument supporting this conjecture is only the consistency that GENERIC provides for combining the time reversible
and nondissipative mechanics with a time irreversible and dissipative mechanism in which  unimportant details are disappearing.

Our discussion of the rate thermodynamics remained  so far in the space $\mathfrak{X}^*(M^{\uparrow})$ in which we did not
relate its elements $X$ to $M^{\uparrow}$. We have so far no connection  between the time evolution of $X\in \mathfrak{X}^*(M^{\uparrow})$
 and the time evolution of  $x\in M^{\uparrow}$.
These two time evolutions become related  if   $X\in \mathfrak{X}^*(M^{\uparrow})$  becomes related to
$x\in M^{\uparrow}$. We have to say something about the function $X(x^*)$. We
recall that we have already addressed  this function in (\ref{disspot}) where we collected properties of the dissipation potential.
In particular, the
dissipation potential $\Xi^{\uparrow}$ has been found to depend on $x^*$ through its dependence on the thermodynamic force $X$ and in a way that
all four properties in (\ref{disspot}) are satisfied.

In order to discuss the compatibility of the  rate thermodynamics with the  thermodynamics presented in Section \ref{structure},
we consider  a macroscopic system  on three levels (\ref{chain1})
in the absence of external and internal forces that prevent approach to equilibrium. The difference between (\ref{chain1}) in
Section \ref{trans} and this section is that $\longrightarrow$ in Section \ref{trans} represent reductions and in this section
it represents rate reductions.
The compatibility of the rate passage $\mathfrak{L}
\rightarrow \mathcal{L}$ with the passage $\mathcal{L} \rightarrow $ \textit{equilibrium} requires
\begin{equation}\label{444}
\Sigma^{\downarrow }(Y)=\Xi^{\uparrow}(Y)
\end{equation}
where $Y$ is the co-vector field on the level $\mathcal{L}$.
This compatibility relation implies then the relation
\begin{equation}\label{enpro}
\mathfrak{S}^{\uparrow}=<S^{\uparrow}_y(y),\dot{y}>=
[<y^*, \Xi^{\uparrow}_{y^*}>]_{y^*=S^{\uparrow}_y(y)}=a[<Y(y^*),\Sigma^{\downarrow}_{Y(y^*)}>]_{y^*=S^{\uparrow}_y(y)}
\end{equation}
between the reduced rate entropy and the entropy production, both on the level $\mathcal{L}$.

Results of the classical nonequilibrium thermodynamics that  gave birth to  the rate thermodynamics
 are recalled in Section \ref{fmt}.

\section{Particular Realizations of GENERIC}\label{realizations}

The unified formulation of equilibrium thermodynamics, non-equilibrium thermodynamics, equilibrium statistical mechanics, and non-equilibrium
statistical mechanics provided by multiscale thermodynamics
has emerged as a collection of common features extracted from  a large body of investigations of macroscopic systems on many different
levels ranging from the equilibrium  to the microscopic.  We now recall some of the principal results on which the multiscale thermodynamics
stands. The feedback of the abstract formulation to the investigation of some specific problems arising in hierarchy reformulations of dynamics
is explored  in Section \ref{hierarchy}.

The first step towards a unified viewpoint of microscopic and mesoscopic dynamics was made by Alfred Clebsch \cite{Clebsch} who cast the
Euler hydrodynamics (i.e. a continuum version of Newton's mechanics) into the Hamiltonian form. In particular  in  Arnold's \cite{ArnoldHam}
formulation,  the Hamiltonian fluid mechanics inspired  efforts to see   also other  mesoscopic nondissipative dynamical theories
(including for instance
kinetic theories) as particular realizations of an abstract Hamilton dynamics.
A modification needed to include dissipative mesoscopic dynamics was made  in \cite{DzVol} and later
in \cite{Grm}, \cite{Morrison}, \cite{Kaufman}. An importance of such  unified formulation has been gradually
realized in \cite{GrmPhysD}, \cite{Morrisonmetriplectic}. Its  usefulness, for instance  in fluid mechanics of complex fluids,  has been first
demonstrated in \cite{GrmPla}, \cite{Berris}. An important step in further theoretical development and in applications was made
in \cite{GrGeneric1}, \cite{GrGeneric2} (where  the acronym GENERIC,
(General Equation for Nonlinear  Equilibrium Reversible-Irreversible Coupling,
appeared for the first time) and also in \cite{Ott}, \cite{Grpost}.
The contact geometry formulation of GENERIC was introduced in \cite{Grmelacont}. A recent systematic presentation of the
multiscale thermodynamics, together with many applications,  can be found in \cite{PKGbook}.

\subsection{Boltzmann Kinetic Equation: Time-evolution Passage}\label{Boltzmann}

Historically the first investigation of the time-evolution passage $\mathcal{L} \rightarrow $ \textit{equilibrium} was made by
Boltzmann \cite{Boltzmann}. The physical system in his analysis   is the ideal gas and the upper level
$\mathcal{L}$ is the level of kinetic theory in which one particle distribution function $f(\rr,\vv)\in M^{\uparrow}$ plays the role of
the state variable ($\rr$ is the position vector and $\vv$ the momentum of one gas particle). The power and the enormous importance
of Boltzmann's results,  as well as  results obtained by his numerous followers,  is not the  narrow focus on the ideal gas
but the physical insight and the mathematical structure involved in the investigation.

We have introduced in Section \ref{grevol} the notion of the entropy $S^{\uparrow}(x)$ as a quantity that plays in mechanics the role of revealing
overall features of the phase portrait. Following Boltzmann's insight, the entropy arises in the investigation of gas dynamics as follows.
The  events in the gas time evolution  that play the most important role in determining the  overall appearance of the phase portrait
are binary collisions. We  therefore consider the free flow of the gas particles and their binary collisions separately. The former
induces directly the time evolution of the one particle distribution function: $f(\rr,\vv,t)=f_0(\Gamma_{-t}(\rr,\vv))$,
where $f_0(\rr,\vv)$ is the distribution function at $t=0$ and $\Gamma_t(\rr,\vv)$ is the trajectory generated by
$\dot{\rr}=\frac{\vv}{m}; \dot{\vv}=0$. The latter  enters  the vector field on $M^{\uparrow}$ indirectly. Complete trajectories
 of colliding particles  are found first  and then transformed into a gain loss balance type vector
field on $M^{\uparrow}$. The transformation is, allegorically speaking, a "retouche" of the trajectories
of colliding particles in which the details are ignored and
only the energy and the momentum conservations are kept.
The Boltzmann entropy is then  born in an analysis of solutions of the Boltzmann equation in which the time evolution
is generated by a vector field that is a sum of the Hamiltonian free flow part  and the modified collision part.

In mathematical terms, the Boltzmann kinetic equation takes the form (\ref{GEN}) with
\begin{eqnarray}\label{Bpot}
&&E^{\uparrow}(f)=\int d\rr\int d\vv\, f  \frac{v^2}{2m};\,\,\, N^{\uparrow}(f)=\int d\rr\int d\vv f \nonumber \\
&& \{A,B\}=\int d\rr\int d\vv f\left(\frac{\partial A_f}{\partial r_i}\frac{\partial B_f}{\partial v_i}-
\frac{\partial B_f}{\partial r_i}\frac{\partial A_f}{\partial v_i}\right)\nonumber \\
&& \Xi^{\uparrow}(f,f^*)=\int d1\int d2\int d1'\int d2'W(f;1,2,1',2')\left(e^{X}+e^{-X}-2\right)\nonumber \\
&& X=f^*(1)+f^*(2)-f^*(1')-f^*(2')
\end{eqnarray}
where $m$ is the mass of one particle. We use hereafter the summation convention over the repeated indices and  the shorthand
notation $1=(\rr_1,\vv_1); 2=(\rr_2,\vv_2),1'=(\rr'_1,\vv'_1); 2'=(\rr'_2,\vv'_2) $. Two particles enter the collision with coordinates $1$
and $2$ and  leave  it with coordinates $1'$ and $2'$. It is assumed that the particles are point particles and their position
coordinates remain unchanged in the collisions (i.e. $\rr_1=\rr'_1 =\rr_2=\rr'_2)$.
The mechanics of binary collisions is introduced into the formulation of the kinetic equation (\ref{GEN}) with (\ref{Bpot}) in two places,
First, in the dissipation potential $\Xi^{\uparrow}(f,f^*)$ in the following restrictions on the choice of $W$: (i)
$W\neq 0$ only if the energy and momentum are conserved, i.e. if $v_1^2+v_2^2=(v'_1)^2+(v'_2)^2$
and $\vv_1+\vv_2=\vv'_1+\vv'_2$, (ii) $W>0$, (iii)
$W$ is symmetric with respect to $1\leftrightarrows 2$ and $(1,2)\leftrightarrows (1',2')$. The second place where the mechanics of binary
collision enters is in the specification of the entropy
$S^{\uparrow}(f)$ that enters  the dissipation potential $\Xi^{\uparrow}(f,f^*)$ in the relation
between $f$  and $f^*$ (i.e.  $f^*=S^{\uparrow}_f$). The Boltzmann entropy
$S^{\uparrow}(f)=-\int d\rr\int d\vv f(\rr,\vv)\ln f(\rr,\vv)$ emerges when the form of the collision gain loss balance calculated from the
collision mechanics (see e.g. \cite{Chapman-Enskog}) is cast into the form
$\Xi^{\uparrow}_{f^*}(f,f^*)$ (the second term on the right hand side of (\ref{GEN}) with $\Xi^{\uparrow}$ given in (\ref{Bpot})).

The form of the dissipation potential $\Xi^{\uparrow}(f,f^*)$ of the collision part of the Boltzmann kinetic equation arises
naturally if we regard
binary collisions as  chemical reactions \cite{Wald}, \cite{Grchemr} in which  two species labeled  by $\vv_1$ and $\vv_2$
react and produce two species labeled by $\vv'_1$ and $\vv'_2$ and vice versa,    The thermodynamic force $X$ is called in chemical
kinetics a chemical affinity.
The dissipation potential $\Xi^{\uparrow}(f,f^*)$ appearing in (\ref{Bpot}) is indeed the dissipation potential arising in chemical
kinetics \cite{Grchemr} (see more in Section \ref{chemkin}.  The property (iv) in (\ref{disspot})
is a straightforward consequence of the symmetries of $W(f;1,2,1',2')$. The coefficient
$a$ appearing in the property (iv) is in this example  $a=1/4$.

We now recall some important properties of solutions to the Boltzmann kinetic equation. We begin with the global existence of its solutions
that has been
proven in     \cite{Leray}. DiPerna and Lions  received for this work  Fields Medal.
Another Fields Medal was received by Cedric Villani \cite{Villani} for  proving  the approach of solutions  of
the Boltzmann kinetic equation to the equilibrium states.

The dissipation equilibrium manifold $\mathcal{M}^{\uparrow (deq)}$ (see Section \ref{GENsection}) is composed of solutions
to $\Xi^{\uparrow}_{f^*}(f,f^*)=0$. i.e. solutions to $X(f^*)=0$. With the Boltzmann entropy, the solutions
are the local Maxwell  distribution functions which are also solutions to
\begin{eqnarray}\label{lMaxw}
&&\Phi^{\uparrow (loc)}_{f(\rr,\vv)}=0;\nonumber \\
&&\Phi^{\uparrow (loc)}(f;e^*(\rr),\uu^*(\rr),n^*(\rr))=-S^{\uparrow}(f)\nonumber \\
&&+\int d\rr e^*(\rr)e(f;\rr) + \int d\rr u^*_i(\rr)u_i(f;\rr) +\int d\rr n^*(\rr)n(f;\rr);\nonumber \\
&&e(f;\rr)=\int d\vv f\frac{\vv^2}{2m};\,\,\,\,\uu(f;\rr)=\int d\vv f\vv;\,\,\,\,n(f;\rr)=\int d\vv f
\end{eqnarray}

The equilibrium manifold $\mathcal{M}^{\uparrow (eq)}$ is composed of solutions to the Boltzmann kinetic equation reached as $t\rightarrow\infty$.
These distribution functions are Maxwell distribution functions that are solutions to
\begin{eqnarray}\label{Maxw}
&&\Phi^{\uparrow}_{f(\rr,\vv)}=0\nonumber \\
&&\Phi^{\uparrow}(f;E^*.N^*)=-S^{\uparrow}(f)+E^*E^{\uparrow}(f)+N^*N^{\uparrow}(f)
\end{eqnarray}
The two manifolds $\mathcal{M}^{\uparrow (eq)}$ and $\mathcal{M}^{\uparrow (deq)}$ are related by
$\mathcal{M}^{\uparrow (eq)}\subset\mathcal{M}^{\uparrow (deq)}\subset M^{\uparrow}$.

The fact that the Boltzmann kinetic equation is a particular realization (\ref{Bpot}) of the abstract GENERIC equation (\ref{GEN})
implies that its solutions approach the local Maxwell distribution functions (\ref{lMaxw}). To prove that  they approach a smaller manifold,
 namely the manifold composed the Maxwell distribution functions  expressing equilibrium states (i.e. solutions to (\ref{Maxw})),
requires an  extra effort \cite{Villani}. The Grad-Villani dissipation enhancement
(see Section \ref{GENsection}), needed to narrow down the asymptotically reached manifold,  arises
due to the presence of the free flow in the vector field.

Beside the opportunity to investigate rigorously the approach to the equilibrium level, Boltzmann's kinetic theory provides  also an opportunity
to investigate the approach to a lower level involving the time evolution (i.e. the situation discussed in Section \ref{ll}).
The mapping (\ref{uppeq}) is chosen as follows
\begin{equation}\label{CHE1}
f(\rr,\vv)\mapsto (\rho(\rr),\uu(\rr),e(\rr))=\left(\int d\vv m f(\rr,\vv), \int d\vv \vv f(\rr,\vv), \int d\vv \frac{\vv^2}{2m} f(\rr,\vv)\right)
\end{equation}
The  \textit{l}-manifold $\mathcal{M}^{\uparrow (l)}\subset M^{\uparrow}$ is searched by a perturbation method in which the dissipation
equilibrium manifold (\ref{lMaxw}) serves as its initial approximation \cite{Chapman-Enskog}.
In this initial approximation the Boltzmann kinetic equation turns into
the Euler hydrodynamic equations (i.e. into  the Hamiltonian part of the hydrodynamic equations).

The Chapman-Enskog method thus begins with the dissipation equilibrium manifold (\ref{lMaxw}), the Euler vector field on its tangent space,
 the Boltzmann entropy,
and the local equilibrium reduced thermodynamic relation in the hydrodynamics state space that is implied  (see (\ref{lMaxw})) by the Boltzmann
entropy.
The next step in the Chapman-Enskog method is a deformation of the  dissipation equilibrium manifold (\ref{lMaxw}),
(that we now denote  $\mathcal{M}^{\uparrow (deq 0)}$)  into
$\mathcal{M}^{\uparrow (deq 1)}$ that is required to be  more invariant than $\mathcal{M}^{\uparrow (deq 0)}$). We say that a
manifold $\mathcal{M}\subset M$ is more invariant, with respect to $\mathcal{F}\in\mathfrak{X}(M)$,  than a submanifold $\mathcal{N}\subset M$
if, roughly speaking, the vector field  $[\mathcal{F}]_{\mathcal{M}}$ is sticking out of $T\mathcal{M}$ more than the vector
field $[\mathcal{F}]_{\mathcal{N}}$ is sticking out of $T\mathcal{N}$.
Results of the investigation will  still, of course,  depend on the  precise meaning we  give to  "sticking out more" and "sticking out less"
(see more in \cite{GorbanKarlin}).

After making the first step in the Chapman-Enskog method we obtain an appropriately deformed manifold $\mathcal{M}^{\uparrow (deq 1)}$ with
the Navier-Stokes-Fourier vector field on its tangent space and a
new entropy $S^{(1)}$
(whose maximization provides $\mathcal{M}^{\uparrow (deq 1)}$) and the corresponding to it new reduced thermodynamic relation.

The Navier-Stokes-Fourier vector field is
the vector field \\$[Boltzmann\,\,vector\,\,field]_{\mathcal{M}^{\uparrow (deq 1)}}$ that is appropriately projected on the tangent space
of the manifold $\mathcal{M}^{\uparrow (deq 1)}$
(see more details
in \cite{Grcheming}, \cite{GorbanKarlin}, \cite{red1}, \cite{red2}, \cite{red3}.
The reducing entropy
$S^{\uparrow}(x)$ is obtained as follows.
Let $\mathcal{M}^{\uparrow (l 0)}$ be the initial manifold with which the Chapman-Enskog iterations begin. In the
context of the reduction of kinetic theory to hydrodynamics the manifold $\mathcal{M}^{\uparrow (l 0)}$ is the manifold formed by  local
Maxwell  distributions.
The manifold corresponding to the first Chapman-Enskog approximation is denoted $\mathcal{M}^{\uparrow (l 1)}$.
Let $S^{\uparrow (0)}(x)$, $S^{\uparrow (1)}(x)$ be the entropies corresponding to
$\mathcal{M}^{\uparrow (l 0)}$, $\mathcal{M}^{\uparrow (l 1)}$  in the sense that $\mathcal{M}^{\uparrow (l 0)}$ is formed by solutions to
(\ref{Phimin}) with   $S^{\uparrow (0)}(x)$ and $\mathcal{M}^{\uparrow (l 1)}$  is formed by solutions to (\ref{Phimin})
with         $S^{\uparrow (1)}(x)$. In the context of the reduction of kinetic theory to hydrodynamics $S^{\uparrow (0)}(x)$ is the Boltzmann
entropy.
This type of the Chapman-Enskog sequence of reducing entropies that is induced by the sequence of the Chapman-Enskog reduced vector fields is
 discussed in \cite{Grcheming} \cite{red1},\cite{red2}, \cite{red3}.

An alternative investigation of the reduction \textit{kinetic theory level} $\rightarrow$ \textit{hydrodynamics level}
that begins with the Grad hierarchy formulation of the kinetic equation \cite{Grad} will be discussed in Section \ref{fmk} and
Section \ref{hierarchy}.

Both the Chapman-Enskog and the Grad types of reductions require a complex investigation of solutions of the kinetic equations. If we however
concentrate our attention  only on kinematics  then the reduction from the kinematics of the one particle distribution function
expressed in the Poisson bracket (\ref{Bpot}) to the kinematics of the hydrodynamic fields expressed mathematically in the
Poisson bracket (\ref{fm3}) is completely  straightforward and completely rigorous. The derivation proceeds as follows.
First, we limit the Poisson bracket in (\ref{Bpot}) to functions $A,B$ that depend on $f$ only through
their dependence on
$(\int d\vv f, \int d\vv \eta(f), \int d\vv \vv f)$, where $\int d\rr\int d\vv \eta(f)$ is a Casimir of the Poisson bracket (\ref{Bpot}).
This means that we replace $A_f$ with $A_{\rho}+\eta_f A_s+\vv A_{\vv}$ and similarly
$B_f$ with $B_{\rho}+\eta_f B_s+\vv B_{\vv}$. Straightforward calculations (see \cite{PKGbook}, references cited therein, and \cite{Ogulbrackets})
  lead then from (\ref{Bpot}) to (\ref{fm3}).

\subsection{Gibbs MaxEnt Passage: Gibbs Equilibrium Statistical Mechanics}\label{Gibbs}

The MaxEnt passage $\mathcal{L} \rightarrow  l$,   discussed in
Section \ref{mp},  was made first by Gibbs \cite{Gibbs} for $\mathcal{L}$ being the microscopic level and
\textit{l} the equilibrium level.
The reducing time evolution equation
describing the preparation process for such passage is not a part of the Gibbs analysis. The preparation process is represented only in a few
requirements: the  gradient part of the reducing time evolution
by a reducing entropy that is required to be maximized,
the Hamiltonian part  by constraints in the maximization.
The applicability of the Gibbs reduction   is universal.

In mathematical terms, the upper state variable is  the $n$-particle distribution function $ f(1,...,n)\in M^{\uparrow}$
$n \sim 10^{23}$ is the number of particles.
The Gibbs MaxEnt reduction  starts with the upper reducing thermodynamic relation
\begin{eqnarray}\label{Geq11}
N^{\uparrow}(f)&=&\int d1,...,\int dn  f(1,...,n)                  \nonumber \\
E^{\uparrow}(f)&=&  \int d1,...,\int dn  f(1,...,n) e(1,...,n)               \nonumber \\
S^{\uparrow}(f)&=& -k_B\int d1,...,\int dn  f(1,...,n) \ln f(1,...,n)
\end{eqnarray}
where $k_B$ is the Boltzmann constant,  $e(1,...,n)$ is the energy (Hamiltonian) of $n$ particles, $k_B$ is the Boltzmann constant.
The passage to the equilibrium thermodynamic relation (\ref{lowthr}) is made in the way described in Section \ref{mp}.

We now compare the Gibbs MaxEnt passage to the equilibrium level with the Boltzmann's time-evolution passage (see Section \ref{Boltzmann})
 also to the equilibrium level. Boltzmann
begins with an insight into the appearance of the phase portrait of the reducing time evolution equation.
The crucial role in the emergence of
the equilibrium pattern in the phase portrait is expected to be played by collisions. The part of the vector field generating
the collision trajectories is thus first  "pre-processed" before putting it back to the total vector field. The pre-processing consists of ignoring
details and keeping only the momentum and energy conservations. In this viewpoint, the pre-processed collision vector field takes the form of
a gain-loss  balance known from chemical kinetics.
An investigation of the
time evolution governed by the Boltzmann equation, i.e. by \textit{free flow vector
field} + \textit{pre-processed collision vector field},  reveals that the approach to the equilibrium level is driven by Botzmann's H-function
that we call the Boltzmann entropy. The equilibrium level can be reached by following the time evolution governed by the Boltzmann equation
or alternatively and equivalently by MaxEnt reduction process in which the Boltzmann H-function   is
maximized subjected to the energy and the number of moles constraints.

Gibbs also begins with an insight into the appearance of the phase portrait. However, instead of expressing it in a modification of the  vector field
that generates it, as Boltzmann does,  Gibbs expresses it
directly in the entropy that generates it in the MaxEnt reduction.
Both the Gibbs entropy (that is universal on the microscopic level) and its maximization  (the MaxEnt principle) are postulated.
The microscopic Hamiltonian vector field is represented in the Gibbs MaxEnt reduction only in the constraint of the Gibbs entropy maximization. The
energy is required to remain unchanged in the reduction.
The Gibbs equilibrium pattern  is also often
called "ergodic" with only very vague  reference   to the rigorous mathematical
definition of ergodicity in the theory of dynamical systems on measurable spaces \cite{Walters}. The phase portrait of the ergodic
(in the rigorous mathematical sense) time evolution does possess the Gibbs pattern but the Gibbs MaxEnt reduction
applies to a much larger class of time evolutions.

There is, of course, an enormous difference between the Boltzmann and the Gibbs approaches  to the  passage $\mathcal{L} \rightarrow l$
in the domain of applicability.
While the Gibbs theory is applicable to all macroscopic systems, the Boltzmann theory is applicable only to ideal gases. The pattern that
in the upper-level phase space in the Gibbs theory characterizes the equilibrium level (as well as the entropy generating it in the MaxEnt
reduction) is universal but it is postulated. In the Boltzmann theory the pattern in the upper-level phase portrait characterizing
the equilibrium level is generated by the time evolution governed by the Boltzmann equation but the analysis is made only for ideal gases.
Nevertheless, as we have already pointed
out in the previous section, the  mathematical structure of the Boltzmann equation has inspired and continues to inspire investigations
of the time evolution of macroscopic systems on all levels.

\subsubsection{Gibbs time-evolution passage}\label{Gibbsevol}

An obvious question is of what is the Gibbs time-evolution passage that becomes  the Gibbs MaxEnt passage
if only an initial state and    the final
state reached as $t\rightarrow \infty$ are considered. This question was already asked  in \cite{Grmelajsf}. We  continue to discuss it here.
The kinematics of the N-particle distribution function is expressed mathematically in the Poisson bracket
\begin{equation}\label{Gkin}
\{A,B\}=\int d1...\int dn f\left[\frac{\partial A_f}{\partial r_{\alpha i}}\frac{\partial B_f}{\partial v_{\alpha i}}-
\frac{\partial B_f}{\partial r_{\alpha i}}\frac{\partial A_f}{\partial v_{\alpha i}}\right]
\end{equation}
Its derivation follows completely the derivation of the Poisson bracket for the 1-particle distribution function appearing in (\ref{Bpot}).
The time evolution equation (\ref{Ham1}) corresponding to the bracket (\ref{Gkin}) is the Liouville equation \cite{Liouville},
\cite{Koopman}, \cite{Carleman}
\begin{equation}\label{Geq}
\frac{\partial f}{\partial t}=-\frac{\partial }{\partial r_{\alpha i}}\left(f\frac{\partial E_f}{\partial v_{\alpha i}}\right)+
\frac{\partial }{\partial v_{\alpha i}}\left(f\frac{\partial E_f}{\partial r_{\alpha i}}\right)
\end{equation}
We note that the Liouville equation (\ref{Geq}) is a linear equation independently of the complexity  of interaction
among the particles.
The Liouville lift transforms  the very nonlinear  particle dynamics in the finite dimensional space
with $(1,...,n)$ as its elements into a linear dynamics in the infinite
dimensional space  with $(f(1,...,n)$ as its elements.

Next, we follow Boltzmann and introduce dissipation.
From the physical point of view, we need to identify an event (or events) in which unimportant details are generated. Such events are analogical
to binary collisions in   ideal  gases.
Let such event be identified. The vector field generating it is replaced by an in-and-out balance generated by  mappings
\begin{equation}\label{Btrans}
\Upsilon(1,...,n)=(1',...,n')
\end{equation}
that we  call a \textit{Boltzmann regularization mappings}. In Boltzmann's analysis of an ideal gas,
the mappings $\Upsilon$ represents transformations of incoming momenta $(\vv_1,\vv_2)$ of a binary collision into the outcoming momenta
$(\vv'_1,\vv'_2)$. The
invariants of  $\Upsilon$ are $ (\rr_1,(\rr_1-\rr_2=0), (\vv_1+\vv_2),(\vv_1)^2 +(\vv_2)^2 )$ expressing  the
physical assumption that the gas particles are point particles and
that the particle trajectories in the collision are determined by Hamilton's mechanics but their details are  ignored,
only the momentum and the energy conservations are honored. In the general setting (\ref{Btrans}) we assume that the mappings are one-to-one
and that their invariants are
\begin{equation}\label{Binv}
\mathcal{B}=\{b_1(1,...,n),...,b_m(1,...,n)\}
\end{equation}
 where $m$ functions  $(b_1,...,b_m)$, satisfy
 \begin{equation}\label{Binv1}
 b_1(1,...,n)=b_1(\Upsilon(1,...,n)),...,b_m(1,...,n)=b_m(\Upsilon(1,...,n))
 \end{equation}
 Still following
Boltzmann's analysis,  we introduce the thermodynamic forces
 \begin{equation}\label{Bforce}
X(f^*)=f^*(1,...,n)-f^*(1',...,n')
\end{equation}
and the dissipation potential $\Xi^{\uparrow}$. We choose $\Xi^{\uparrow}$ to be the same
as the one appearing in (\ref{Bpot}) but with $X$ given in (\ref{Bforce}). We now add
to the right hand side of the Liouville equation (\ref{Geq})
an additional term  $\Xi_{f^*}$. The resulting equation
\begin{equation}\label{GibbsB}
\frac{\partial f}{\partial t}==\frac{\partial }{\partial r_{\alpha i}}\left(f\frac{\partial E_f}{\partial v_{\alpha i}}\right)+
\frac{\partial }{\partial v_{\alpha i}}\left(f\frac{\partial E_f}{\partial r_{\alpha i}}\right) +\Xi_{f^*}
\end{equation}
possess the GENERIC structure and consequently, see Section \ref{GENsection}, its solutions approach
solutions to $X=0$. Such solutions form a manifold $\mathcal{M}^{\uparrow (eq)}=\{f\in M^{\uparrow}|f^*=\sum_{i=1}^{m}<b_i^*,b_i>\}$,
parametrized by $b_1^*,...,b_m^*$. With the Gibbs entropy, the dissipation potential $\Xi$ given in (\ref{Bpot}, and the thermodynamic force
$X$  (\ref{Bforce}) the time evolution equation (\ref{GibbsB}) becomes
\begin{eqnarray}\label{GibbsBB}
\frac{\partial f}{\partial t}&=&-\frac{\partial }{\partial r_{\alpha i}}\left(f\frac{\partial E_f}{\partial v_{\alpha i}}\right)+
\frac{\partial }{\partial v_{\alpha i}}\left(f\frac{\partial E_f}{\partial r_{\alpha i}}\right)\nonumber \\
&& + \int d1...\int dn W(1,...,n,1',...,n')(f(1',...,n')-f(1,...,n))
\end{eqnarray}
where $W$ is symmetric with respect to $(1,...,n)\rightarrow (1',...,n')$, $W\geq 0$, and $W=0$ unless  (\ref{Binv1}) holds.
As it is the case with the Liouville equation (\ref{Geq}), the the complex and typically very nonlinear transformations $\Upsilon$
in the Boltzmann
regularization mappings turn in the Liouville lift into a linear collision-like term.

The Boltzmann-inspired "retouche" of the phase portrait that we presented above is similar to the Ehrenfest regularization
 (Ehrenfest "retouche") \cite{Ehrenfest},
\cite{red2} in which very small pieces of  trajectories are pre-processed.

A likely scenario of the Gibbs time-evolution passage to the equilibrium level is the following. The time evolution begins with a weak
dissipation, i.e. with a  large set (\ref{Binv}) of invariants which means that only a few details are being ignored.
In the course of the time evolution the dissipation increases due to the
Grad-Villani enhancement (see Section \ref{GENsection}) until the set $\mathcal{B}$ of invariants of the Boltzmann regularization mappings
(\ref{Btrans}) becomes the small set $\{f\in M^{\uparrow}|E(f)=E;  N(f)=N\}$. In order to continue to discuss this
scenario one must discuss in particular
the set of the invariants (\ref{Binv}) representing the seed of the dissipation,  applicability of the Grad-Villani enhancement
of the dissipation, and the
entropy entering  the relation between $f$ and $f^*$. All these discussions have to arise in an analysis of solutions to (\ref{Geq}).

\subsection{Euler Fluid Mechanics: Local Conservation Laws}\label{lcl}

The level of fluid mechanics is the oldest \cite{Euler} and undoubtedly the most important (at least from the application point of view)
mesoscopic level. It has also served as a nucleus of other nearby levels, like for instance the
level of mechanics of solids,
the level of the mechanics of complex fluids (rheology), the classical nonequilibrium thermodynamics, and also  many fields
in mathematics.
We recall below some aspects of its relations to  mechanics, kinetic theory, and equilibrium thermodynamics.  We also recall some branches
of physics and mathematics that grew out of these investigations.

\subsubsection{Relation to Newton's mechanics}\label{fmm}

Leonhard Euler \cite{Euler} has introduced fluid mechanics as a continuum version of Newton's mechanics of particles.
The state variables are the fields
\begin{equation}\label{fm10}
(\rho(\rr), e(\rr),\uu(\rr)).
\end{equation}
of mass, energy, and momentum respectively.
 The total mass
$M^{(fm)}=\int d\rr \rho(\rr)$, the total energy $E^{(fm)}=\int d\rr e(\rr)$ and the total momentum $\UU^{(fm)}=\int d\rr \uu(\rr)$
remain unchanged during the time evolution. If we limit ourselves to fluids with only local interactions then
this property implies that the time evolution
equations form a system of local conservation laws (also called balance laws)
\begin{eqnarray}\label{loccons}
\frac{\partial \rho}{\partial t}&=&-\frac{\partial J_i^{(\rho)}}{\partial r_i}\nonumber \\
\frac{\partial e}{\partial t}&=&-\frac{\partial J_{i}^{(e)}}{\partial r_i}\nonumber \\
\frac{\partial u_i}{\partial t}&=& -\frac{\partial J_{ij}^{(u)}}{\partial r_j}
\end{eqnarray}
The fields
$(\JJ^{(\rho)},\JJ^{(e)},\JJ^{(u)})$ appearing in (\ref{loccons}) are fluxes. Their specification
as functions of the state variables
$(\rho(\rr),e(\rr),\uu(\rr))$ is
called a \textit{constitutive relation} (see e.g. \cite{Truesdell}, \cite{Hungary}).
The individual nature of the fluids is expressed in (\ref{loccons}) in the constitutive relations.
The third equation in (\ref{loccons}) has two physical interpretations, one as  a local conservation law (momentum conservation), and the other
as a continuum version of Newton's law (mass times acceleration equals force).

The  Hamilton formulation of the governing equations of fluid mechanics has appeared
in 1859  in Ref.\cite{Clebsch}.  We present it in the form introduced by    Arnold \cite{ArnoldHam}. We begin with only the field $\uu(\rr)$
in the set of  the state variable (\ref{fm10}).
Our objective is to find a particular realization of (\ref{Ham1}) with $x=\uu(\rr)$.
In order to find  the kinematics of $\uu(\rr)$ (i.e. in order to  determining the Poisson bivector $L^{\uparrow}$)  we turn to the
physics of continuum.  Following  Euler \cite{Euler}, continuum is the space $\mathbb{R}^3$ and its motion
is a Lie group of transformations $\mathbb{R}^3\rightarrow \mathbb{R}^3$. Arnold \cite{ArnoldHam} has realized  that the
momentum field $\uu(\rr)$ is an element of the dual of the Lie algebra that is associated with the Lie group of the transformations
$\mathbb{R}^3\rightarrow \mathbb{R}^3$ and consequently that
the Poisson bracket that is  canonically associated with the Lie algebra \cite{Marsden}, \cite{Marsden1}
 (that in the case of Lie group of  transformations
$\mathbb{R}^3\rightarrow \mathbb{R}^3$) has the form
$\{A,B\}=\int d\rr u_i\left(\frac{\partial A_{u_i}}{\partial r_j}B_{u_j}-\frac{\partial B_{u_i}}{\partial r_j}A_{u_j}\right)$)
expresses mathematically the kinematics of the continuum.

In order to identify the kinematics of  the full set (\ref{fm10}) of the state variables that also satisfies the degeneracy requirement
(see Section \ref{HHam}),
we make an extra hypothesis about
the time evolution on the level of fluid mechanics. We replace the energy field $e(\rr)$ in (\ref{fm10}) with another scalar
field $s(\rr)=s(\rho,e,\uu;\rr)$ that is required to satisfy:
\begin{eqnarray}\label{fm1}
&& (\rho(\rr),e(\rr),\uu(\rr))\leftrightarrows (\rho(\rr),s(\rr),\uu(\rr))\,\, is\,\,a\,\,one-to-one\,\,transformation\nonumber \\
&& \frac{\partial s}{\partial t}=s_{\rho}\frac{\partial \rho}{\partial t}+s_e\frac{\partial e}{\partial t}
+s_{u_i}\frac{\partial u_i}{\partial t}=-\frac{\partial J_{i}^{(s)}}{\partial r_i}
\end{eqnarray}
where $\JJ^{(s)}$  is a flux of the field $s$. The flux $\JJ^{(s)}$  is a function of the hydrodynamic fields. Its specification is a part of
the constitutive relation.  The physical interpretation
of (\ref{fm1}) will appear   in Section \ref{fmt} in the discussion of the relation of fluid mechanics with equilibrium thermodynamics.
In the rest of this section we shall use the fields $(\rho(\rr),s(\rr),\uu(\rr))$ as the state variables of fluid mechanics.

The two scalar fields $(\rho(\rr), s(\rr))$
are assumed to be passively advected with the motion of the continuum. With the use of the concept of semi-direct product
\cite{Marsden},\cite{Marsden1}
the complete
Poisson bracket expressing the kinematics of  $(\rho(\rr),s(\rr),\uu(\rr))$ is given by
\begin{eqnarray}\label{fm3}
\{A,B\}&=& \int d\rr \left[ u_i\left(\frac{\partial A_{u_i}}{\partial r_j}B_{u_j}-\frac{\partial B_{u_i}}{\partial r_j}A_{u_j}\right)
\right.\nonumber \\
&& \left.+ \rho\left(\frac{\partial A_{\rho}}{\partial r_j}B_{u_j}-\frac{\partial B_{\rho}}{\partial r_j}A_{u_j}\right)\right.\nonumber \\
&&\left. s\left(\frac{\partial A_{s}}{\partial r_j}B_{u_j}-\frac{\partial B_{s}}{\partial r_j}A_{u_j}\right)\right]
\end{eqnarray}

The equations (\ref{Ham1}) governing the Hamiltonian time evolution of $(\rho(\rr),s(\rr),\uu(\rr))$  are thus
\begin{eqnarray}\label{lloccons}
\frac{\partial \rho}{\partial t}&=&-\frac{\partial J_i^{(\rho)}}{\partial r_i}\nonumber \\
\frac{\partial s}{\partial t}&=&-\frac{\partial J_{i}^{(s)}}{\partial r_i}\nonumber \\
\frac{\partial u_i}{\partial t}&=& -\frac{\partial J_{ij}^{(u)}}{\partial r_j}
\end{eqnarray}
where
\begin{eqnarray}\label{fm11}
J^{(\rho)}_i&=&\rho E^{\uparrow}_{u_i}\nonumber \\
J^{(s)}_i&=&s E^{\uparrow}_{u_i}\nonumber \\
J^{(u)}_{ij}&=&u_i E^{\uparrow}_{u_j}+p\delta_{ij}\nonumber \\
p&=& -e+\rho E^{\uparrow}_{\rho}+sE^{\uparrow}_s+u_i E^{\uparrow}_{u_i}
\end{eqnarray}
We see thus that the requirement expressed in the second equation in (\ref{fm1}) is satisfied and thus $S=\int d\rr s(\rr)$ is the Casimir of
the Poisson bracket (\ref{fm3}).

The Hamiltonian formulation of Euler's equations (\ref{lloccons}) has at least four advantages: (i) the constitutive relation  for the
nondissipative part of the time evolution are specified (see (\ref{fm11}))
with only the energy $E(\rho,s,\uu)$ remaining to be determined, (ii) it provides a framework for investigations dynamics of
more general fluids (e.g. complex fluids
studied in rheology \cite{Grcheming})  for which
the framework (\ref{loccons}) of balance laws cannot be used, (iii) it can also be used on other mesoscopic
and microscopic levels of description,  (iv) it offers promising new approaches to numerical fluid mechanics \cite{nummech}.
 In spite of these obvious advantages the Hamiltonian formulation is still
absent in  most standard textbooks of fluid mechanics.

The gradient part of the time evolution will appear  in Section \ref{fmt} in the discussion of the relation between the level of fluid mechanics and
the equilibrium level.

Looking at (\ref{loccons}) and  (\ref{fm1}) just from the mathematical point of view, we see a system of local conservation laws (\ref{loccons})
implying  another companion conservation law (\ref{fm1}). Are there any   mathematical consequences of the physical regularity of
(\ref{loccons}) expressed in the requirement  (\ref{fm1})?
Godunov \cite{Godunov}, \cite{Godunov1}, (see also \cite{FL}, \cite{BR}, \cite{MullerRugg}, \cite{Peshkov})
  have shown that the physical regularity implies the mathematical
regularity  in the sense that (\ref{fm1}) guarantees that the  Riemannian problem for (\ref{lloccons}) is well posed.
 More specifically, (\ref{fm1}) implies that
(\ref{loccons}) rewritten in the conjugate state variables is a system of symmetric local conservation laws.

The observation that (\ref{loccons}), beside being the system of local conservation laws,  is also a Hamiltonian system,  and that (\ref{fm1})
is a stronger formulation of  the  degeneracy of the Hamiltonian structure  ($\int d\rr s(\rr)$ is a Casimir) evokes several questions that
remain unanswered.
For instance:  (i) when does a general system of local conservation laws possess the Hamiltonian structure, (ii) does the degeneracy of
a Hamiltonian system implies an  increase in  its mathematical regularity.

 \subsubsection{Relation to more microscopic levels}\label{fmk}

The level of fluid mechanics is presented  in the previous section as a continuum version of Newton's (or Hamilton's ) dynamics. Let us now take
an upper mesoscopic level $\mathcal{L}$ that involves more details than the level of fluid mechanics
(e.g. the level of kinetic theory) and consider the passage
$\mathcal{L} \rightarrow$ \textit{fluid mechanics}.  We have already recalled one   such passage
with $\mathcal{L}$ being the level of kinetic theory in Section \ref{fmk}. An alternative way (a way based on the hierarchy formulation of
the Boltzmann equation) to make the same passage is  discussed below in Section \ref{hierarchy}.

With the microscopic level (i.e. a level on which $n$-particle distribution function,  $n\sim 10^{23}$,  rather than 1-particle distribution function
serves as  the state variable)
playing the role of the upper level $\mathcal{L}$, the passage $\mathcal{L} \rightarrow $ \textit{fluid mechanics}
was investigated by Kirkwood \cite{Kirkwood}. This type of investigations has led to the theoretical fluid mechanics
of complex fluids as for example polymeric fluids and  suspensions \cite{Bird}.

We make two remarks. First, we  note an important difference between the multiscale viewpoint of the passage
\textit{Boltzmann kinetic equation} $\rightarrow$ \textit{fluid mechanics} and its classical analysis  found  for example in Refs.
\cite{Chapman-Enskog}, \cite{deGrootM}, \cite{Ruggeri1}, \cite{Wenan}, \cite{Struch}, \cite{EIT}. In the latter the Boltzmann kinetic equation
plays the role of a microscopic basis for the classical nonequilibrium thermodynamics. In the former, the Boltzmann
kinetic theory as well as the classical nonequilibrium thermodynamics are two particular realizations, on two different levels,
of a single but  abstract nonequilibrium thermodynamics.

In the second remark we note  an obvious paradox in the  investigation of \textit{Boltzmann kinetic equation} $\rightarrow$ \textit{fluid mechanics}.
 The Boltzmann kinetic theory is applicable only to ideal gases while the domain of applicability of fluid mechanics includes a large family of fluids.
The usefulness of the investigations \textit{Boltzmann kinetic equation}$\rightarrow$ \textit{fluid mechanics}
is an indirect proof of the usefulness of seeing
mesoscopic  dynamical systems   in a modular way as it is done for example in Section \ref{te}. What  transpires  from
kinetic theory to fluid mechanics are only some of its modules (in particular the overall mathematical structure), not the complete theory
(in particular not specific energies and specific entropies). The completely straightforward and completely rigorous derivation of
the Poisson bracket expressing kinematics of the hydrodynamic state variables from the Poisson bracket expressing kinematics of
one particle distribution function, that we recalled at the end of Section \ref{Boltzmann},  illustrates well this point.
As we shall see also in the next section, some modules of the mathematical structure of the Boltzmann kinetic equation that are revealed in
its Grad hierarchy formulation have inspired, and continues to inspire, not only the classical
fluid mechanics but also its extensions towards dense fluids, polymeric fluid, and many other types of complex fluids.

 \subsubsection{Relation to equilibrium thermodynamics: nonequilibrium thermodynamics}\label{fmt}

Inquiries into  relations between fluid mechanics and equilibrium thermodynamics gave rise to  nonequilibrium thermodynamics.
Our objective in this section is  to identify  in its   tumultuous  history
a path pointing   to the multiscale thermodynamics discussed in this paper. We present the path  as a sequence of four steps.

The base on which the classical nonequilibrium thermodynamics stands is the continuum mechanics introduced by
Euler and Bernoulli \cite{Euler}. If we put it into the context of Section \ref{fmm}, it is
the fluid mechanics with only the momentum field $\uu(\rr)$ playing the role of the state variable. In other words, the fluids under investigation are
isothermal and incompressible. This type of fluid mechanics has played  and continues to play an enormously important role
in all types of the most basic as well as the most advanced technologies.

The first step towards the multiscale thermodynamics is mechanics  of non isothermal and compressible fluids and investigations of its
compatibility with the classical equilibrium thermodynamics.
Two extra fields, namely the fields of mass density and
internal energy are adopted to the set of state variables. In such enlarged setting the fluid mechanics becomes essentially a local classical
thermodynamics superimposed on the mechanics of continuum. The equilibrium fundamental thermodynamic relation becomes a local equilibrium
fundamental thermodynamic relation, the entropy conservation takes the form of the local conservation law  (\ref{fm1}). The Navier-Stokes friction
and the Fourier  heat diffusion  enter the entropy production (or the dissipation potential)  that with   the entropy  are two potentials
of non mechanical origin that join the formulation of  fluid mechanics.

Combinations of mechanics and  equilibrium thermodynamics inspired  also  more abstract viewpoints. Their explorations
 constitute the second step in the evolution path of the nonequilibrium thermodynamics.
The first example  of an abstraction inspired by fluid mechanics is the replacement of (\ref{loccons}) with a general
system of local conservation laws governing the time evolution of $n$ fields $(\xi_1(\rr),...,\xi_n(\rr))$
with an extra companion local conservation law governing the time evolution of $(n+1)$th field $\xi_{n+1}(\rr)$ that is a convex function
of $(\xi_1(\rr),...,\xi_n(\rr))$. From the physical point of view, the $(n+1)$th field is the entropy field (see
the end of Section \ref{fmm}).

The second example of the abstraction is the emergence (already in early stages of the development of
nonequilibrium thermodynamics \cite{Prigogine}, \cite{Onsager2}, \cite{Gyarmati})
 of  the   concepts of  entropy production, thermodynamic forces, and thermodynamic fluxes \cite{Brescia}.
Their particular realizations in the context of fluid mechanics
have served as their illustrations but they were seen from the beginning  as abstract concepts. The thermodynamic fluxes and
thermodynamic forces  together form the entropy production
\begin{equation}\label{entpr3}
\mathfrak{S}=<X,J>
\end{equation}
or in terms of the dissipation potential $\Xi(X)$
\begin{equation}\label{entpr4}
\mathfrak{S}=<X,\Xi_X>
\end{equation}
Particularly significant  in  this line of research are results of  Onsager \cite{Onsager}
who showed that in the case of the
quadratic dissipation potential
$\Xi=<X,\Lambda X>$ (that physically corresponds to situations with small $X$ and thus situations when the macroscopic systems
under investigation are close to equilibrium) the operator  $\Lambda$ is symmetric and positive definite.
The rate thermodynamics that we recalled in Section \ref{rt} is  an incorporation  of this type of investigations into the larger context of
multiscale thermodynamics.

The third example is  Truesdell's  axiomatic formulation of continuum mechanics \cite{Truesdell}. While his choice of  axioms may be questioned,
the emphasize on the abstract mathematics is unquestionably a significant contribution to fluid mechanics. For instance restrictions on the
choice of constitutive relations brought about by the requirement of the  entropy increase have been first investigated in Truesdell's
formulation of fluid mechanics \cite{Coleman}.

In the spirit of multiscale thermodynamics discussed in this paper, an important criterion
for abstract formulations is the
occurrence and applicability on all levels. Not all Truesdell's axioms   fulfill this criterion. For instance the local temperature cannot
be seen on more microscopic levels as a fundamental state variable (see  Section \ref{mp}).

The third step on the path to the multiscale thermodynamics is seeing the  Boltzmann
kinetic theory as nonequilibrium thermodynamics itself, not only as
a microscopic basis for the classical (i.e. fluid-mechanics-based)  nonequilibrium
thermodynamics.

The fourth step on the path to the multiscale thermodynamics is the necessity to enlarge the set of the  five hydrodynamic fields
playing the role of state variables in the classical fluid mechanics when dealing with complex fluids (as for example polymeric fluids
and suspensions). The molecules (or alternatively particles in suspensions) composing the complex fluids deform and reorient themselves at the same time scale as the
hydrodynamic fields evolve. Consequently, extra fields characterizing the internal structure have to be adopted to the set of state variables.
But then the system of local conservation laws (also called "balance laws") (\ref{loccons}) cannot be the point of departure (as it is
in the classical fluid mechanics and the classical nonequilibrium thermodynamics) since the extra fields are typically not conserved.
What is then an overall structure that would replace (\ref{loccons})? In the setting of mesoscopic thermodynamics, the answer is: it is
the Hamiltonian (or the GENERIC) structure.

\subsection{Guldberg-Waage Chemical Kinetics: Dissipation Potentials}\label{chemkin}

Historically  the first investigation of the  time-evolution passage, that we recalled in  Section \ref{Boltzmann},  is  also historically the
first introduction of the  dissipation-potential gradient dynamics (\ref{gr2}). Initially, the Boltzmann
collision term did not have the form of the right hand side of (\ref{gr2}). Its dissipation-potential formulation
became possible  \cite{Grm} only after
an observation (made by Ludwig Waldmann in \cite{Wald}) that binary collision can be seen
as  chemical reactions and after the  gradual realization
\cite{deD}, \cite{Feinberg}, \cite{Sien}, \cite{Yab}, \cite{*gengrad}  that the time evolution arising in chemical kinetics can be
cast into the form (\ref{gr2}).
In this section we  argue   that the chemical kinetics provides a particulary
suitable setting for a deeper investigation of solution to the GENERIC equation.

A need to extend  linear relations between thermodynamic fluxes and thermodynamic forces to  nonlinear relations appeared
very clearly  in particular  in chemical kinetics
\cite{GW} describing the time evolution of chemically reacting species. Let us consider $q$ chemical reactions among $p$ components
(we assume $p>q$)
\begin{equation}\label{chemreac}
\alpha_{1j}\mathbb{A}_i+...+\alpha_{pj}\mathbb{A}_p\leftrightarrows \beta_{1j}\mathbb{A}_i+...+\beta_{pj}\mathbb{A}_p
\end{equation}
where $\mathbb{A}_1,...,\mathbb{A}_p$ denote the species. The quantities
\begin{equation}\label{gam}
\gamma_{ij}=\alpha_{ij}-\beta_{ij};\,\,\,i=1,...,p;\,\,\,j=1,...,q
\end{equation}
are called stoichiometric coefficients
and the matrix $\Gamma=(\gamma_{ij}); \,\,\,i=1,...,p;\,\,\,j=1,...,q$ is called a stoichiometric matrix.

The state variables in isothermal chemical kinetics are $\nn=(n_1,...,n_p)$ denoting the number of moles of the species, and the
constant temperature $T$.
The equations
\begin{equation}\label{cheq1}
\dot{n}_i=\gamma_{ij}J_j
\end{equation}
governing the time evolution of $\nn$ resemble  the system of local conservation laws (\ref{loccons}). The gradient $\nabla$ appearing in
(\ref{loccons}) is replaced by the stoichiometric matrix $\Gamma$ and  the fluxes $(J_1,...,J_q)$ in (\ref{cheq1}) are extends of the reactions.

Guldberg and Waage \cite{GW} have completed (\ref{cheq1}) with  the mass-action-law constitutive relations
\begin{equation}\label{gw1}
J_j=\overrightarrow{k}_jn_1^{\alpha_{1j}}...n_p^{\alpha_{pj}}-\overleftarrow{k}_jn_1^{\beta_{1j}}...n_p^{\beta_{pj}};\,\,\,j=1,...,q
\end{equation}
where $\overrightarrow{k}_j, \overleftarrow{k}_j$ are rate coefficients of the forward $j-th$ reaction and the backward $j-th$ reaction
respectively.

It has been gradually \cite{deD}, \cite{Feinberg}, \cite{*gengrad}, \cite{Sien}, \cite{Yab}, \cite{GGG}, \cite{Grchemr}
realized that (\ref{cheq1}) with
the constitutive relation (\ref{gw1}) can be cast into the form
\begin{equation}\label{chemm}
\dot{n}_i=-\Xi_{n_i^*}(\nn,T,\XX)
\end{equation}
with the dissipation potential $\Xi$ given in (\ref{Bpot} in which
the thermodynamic forces $\XX=(x_1,...,X_q)$, called in chemical kinetics  affinities, are
\begin{equation}\label{aff}
X_j=\gamma_{ji} n^*_i; \,\,j=1,...,q
\end{equation}
$n_i^*=\Phi_{n_i}(\nn,T)$, $W$ is expressed in terms of the rate coefficients $\overrightarrow{k}_j, \overleftarrow{k}_j$
(see details in \cite{Grchemr})),  and an
appropriately  chosen   thermodynamic potential  $\Phi(\nn,T)$ (see details in \cite{Grchemr}).

\subsubsection{GENERIC formulation of chemical kinetics}

The state spaces in the Guldberg-Waage dynamics are finite dimensional. This is obviously an advantage in the investigation of solutions.
Another advantage of the Guldberg-Waage dynamics, in particular in the context of
the multiscale thermodynamics, is a natural  appearance  of intermediate levels.
We begin by extending (\ref{chemm})  to a full GENERIC equation (\ref{GEN}), the intermediate levels are discussed in Section \ref{problem} below.

From the physical point of view, the  extension of (\ref{chemm}) to GENERIC  expresses mathematically an inclusion of inertia into chemical reactions.
The fluxes $\JJ=(J_1,...,J_q)$ are promoted to the status of
independent state variables. We are thus making in chemical kinetics the same type of extension as the one made in fluid mechanics in \cite{MullerRugg},
\cite{EIT}.
The time evolution of the fluxes is assumed to be faster than the time evolution of $\nn$. After the fast
time evolution of the fluxes is completed,  the subsequent slow time evolution of $\nn$ is expected to be govern by the standard chemical kinetics
equations (\ref{chemm}).

Equations
\begin{equation}\label{genchem}
\left(\begin{array}{cc}\dot{\nn}\\\dot{\ww}\end{array}\right)=\left(\begin{array}{cc}0&\Gamma \\-\Gamma^T&0\end{array}\right)
\left(\begin{array}{cc}\nn^{\star}\\ \ww^{\star}\end{array}\right) - \left(\begin{array}{cc}0\\ \Theta_{\ww^{\star}}\end{array}\right)
\end{equation}
governing such time evolution of the extended set of state variables have been
introduced in \cite{GGG}, \cite{PKGbook}:
The extra state variables $\ww= (w_1,...,w_q)$ are related to  $\JJ=(J_1,...,J_q)$ by $w_j^{\star}=J_j; j=1,...,q$,
where  $\nn^{\star}=\Phi^{(ext)}_{\nn}$,
$\ww^{\star}=\Phi^{(ext)}_{\ww}$,
$\Phi^{(ext)}(\nn,\ww)$ is the thermodynamic potential in the extended theory, and $\Theta(\nn^{\star},\ww^{\star})$ is the  dissipation potential
in the extended theory.
First, we show that (\ref{genchem}) is a particular
realization of the GENERIC equation {\ref{GEN}) and then we find the relation between the extended dissipation potential $\Theta$
appearing in the extended chemical kinetics (\ref{genchem}) and the
dissipation potential $\Xi$ appearing in the standard chemical kinetics (\ref{chemm}).

The first term on the right hand side of (\ref{genchem}) is Hamiltonian since \\$(A_{\nn},A_{\ww})\left(\begin{array}{cc}0&\Gamma\\-\Gamma^T&0
\end{array}\right)\left(\begin{array}{cc}B_{\nn}\\B_{\ww}\end{array}\right)$ is a Poisson bracket (because the matrix \\
$\left(\begin{array}{cc}0&\Gamma\\-\Gamma^T&0
\end{array}\right)$ is skewsymmetric
and independent of the state variables), $A$ and $B$ are real valued sufficintly regular functions of the state variables. The time evolution
equation (\ref{genchem}) is thus a particular realization of (\ref{GEN}).

Now we turn to the problem of finding the relation between the dissipation potential $\Xi$ appearing in (\ref{chemm})
and the extended dissipation potential $\Theta$ appearing in (\ref{genchem}).
We assume that the time evolution of $\ww$ is faster than the time evolution of $\nn$. Moreover,  we require that when the fast variable $\ww$
reaches its stationary state, i.e. when
\begin{equation}\label{chch1}
-\Gamma^T\nn^{\star}-\Theta_{\ww^{\star}}=0
\end{equation}
then the subsequent time evolution is governed by (\ref{chemm}). This requirement is satisfied provided  the dissipation potentials
$\Xi (X)$ and the extended dissipation potential $\Theta(Y)$ are related by
\begin{equation}\label{chch2}
\Xi_{\nn^*}=\left[\Theta^{\dag}_{\YY^{\dag}}(\YY^{\dag})\right]_{\YY^{\dag}=-\XX}
\end{equation}
and the thermodynamic potential $\Phi(\nn,T)$ and the extended thermodynamic
potential $\Phi^{(ext)}(\nn, \ww,T)$ are related in such a way that $\nn^*=\nn^{\star}$.

The superscript $\dag$
denotes the   conjugation with respect to the dissipation potential $\Theta$, i.e.  $(\ww^{\star})^{\dag}=
\Theta_{\ww^{\star}}$. By the symbol $\YY$ we denote the extended thermodynamic force $\YY=\Gamma \ww^{\star}$.
The dissipation potential $\Theta^{\dag}(\YY^{\dag},T)$ is the Legendre
transformation of $\Theta(\YY,T)$.

We now prove that (\ref{chch2}) implies the compatibility between (\ref{chemm}) and (\ref{genchem}).
We introduce the dissipation thermodynamic potential
 $$\Psi(\YY,T;\YY^{\dag})=-\Theta(\YY,T)+<Y^{\dag},Y>$$,
and recall  that $Y=\Theta^{\dag}_{Y^{\dag}}$. Finally,  we note that (\ref{chch1}) is the equation
\\$\left[\Psi_{\YY}(\YY,T;\YY^{\dag})\right]_{\YY^{\dag}=-\XX}=0$.

Before leaving this section   we make a comment about
the difference between the potentials like  energy, entropy,
 number of moles  and the dissipation potential. Gradients of the former  generate the forces
driving the time evolution,  gradients of  the latter transform  the forces (co-vectors) into  vector fields.
Specifically, the forces generated by the energy are transformed into the Hamiltonian vector fields by the Poisson bracket.
The forces generated by the entropy are transformed into  vector fields  by the dissipation potential. The dissipation potential in
gradient dynamics
is  thus a counterpart of the Poisson bracket in the Hamiltonian dynamics.
Locally,  in a small neighborhood of $\XX=0$,  the dissipation potentials become quadratic. Such
quadratic dissipation potentials can be then interpreted as a dissipation brackets (introduced in \cite{Grm}). In this linearized
formulation there are  thus two brackets transforming forces into vectors. The symmetric dissipation bracket can be extended to a nonlinear
dissipation potential. No such extension can be made for the skewsymmetric Poisson bracket.

If both the Poisson bracket and the dissipation potentials
participate in dynamics  then the Poisson bracket is completely insensitive
to the forces generated by the entropy and the total number of moles. The  dissipation potential is on the other hand
completely insensitive to the forces
generated by the energy and the total number of moles. The insensitivities are  mathematically expressed in degeneracies of the Poisson
brackets and the dissipation potentials.

In the contact structure formulation of the GENERIC dynamics (see Section \ref{CS}) both the
Poisson bracket and the dissipation potential become potentials generating the contact-structure preserving time evolution.
The potentials like the energy, the entropy and number of moles determine the Legendre submanifold on which the time evolution takes place.
Dissipation potentials arise  \cite{Mielke2} also in the stochastic
thermodynamics in  extensions to the large deviation  stochastic theory.

\subsubsection{Problem: Grad-Villani dissipation enhancement in chemical kinetics}\label{problem}

Let some of the chemical reactions (\ref{chemreac})
be faster than the others. Specifically, let the first $m$ reactions be fast
and the remaining $q-m$ reactions  slow. We  write  the
stoichiometric matrix in the form
\begin{equation}\label{fastslow1}
\Gamma =\Gamma^{(fast)}+\Gamma^{(slow)}
\end{equation}
The matrix $\Gamma^{(fast)}=(\gamma^{(fast)}_{ij}=\gamma_{ij})$, $i=1,...,p$ ; $j=1,...,m$ and
$(\gamma^{(fast)}=0)$, $i=1,...,p$;   $j=m+1,...,q$. The matrix
$\Gamma^{(slow)}=(\gamma^{(slow)}_{ij}=\gamma_{ij})$  $i=1,...,p$, $j=m+1,...,q$ and $(\gamma^{(slow)}=0)$, $i=1,...,p$;   $j=1,...,m$.
The split (\ref{fastslow1}) induces (see (\ref{aff}))  the splits
$\XX=\XX^{(fast)}+\XX^{(slow)}$ and $\YY=\YY^{(fast)}+\YY^{(slow)}$
The passage to the chemical equilibrium at which $\XX=0$ can be seen as a two stage process. The first stage is
the approach to the fast equilibrium at which $\XX^{(fast)}=0$. The fast approach is then followed by a slow time evolution that terminates
 at the total
equilibrium at which both $\XX^{(fast)}$ and $\XX^{(slow)}$ equal zero. This type of the time evolution has been recently investigated in the
setting of the stochastic thermodynamics in \cite{Mielke}. With the results obtained in the preceding section we can investigate the same
problem in the setting of  multiscale thermodynamics. In this paper we limit ourselves only to formulating  two questions.

\textit{Question 1}

How exactly are   solutions to (\ref{chemm}) and solutions to (\ref{genchem}) related?  (The dissipation potential
$\Theta$  appearing in (\ref{genchem}) is assumed to satisfy (\ref{chch2}))

\textit{Question 2}

We modify Eq.(\ref{genchem}) by replacing
 the dissipation potential $\Theta$ with the fast dissipation potential $\Theta^{(fast)}=\left[\Theta\right]_{\YY=\YY^{(fast)}}$
 that provides  a weaker dissipation. Do
solutions to such modified equation with a weaker dissipation still approach the same chemical equilibrium state as solutions to (\ref{genchem})?
Investigations of this question are  investigations of the Grad-Villani dissipation enhancement (see Section \ref{GENsection}) in the context of
chemical kinetics.

\subsection{Statistical Mechanics}\label{statmech}

The problems discussed in this and the previous sections belong to statistical mechanics. The way we presented and discussed
microscopic, mesoscopic, and macroscopic dynamics  suggests
an alternative  view of statistical mechanics. First we recall the conventional  viewpoint  
and then  we introduce its alternative.

Traditionally, statistical mechanics is divided  into two parts: equilibrium and nonequilibrium.
In the  broader viewpoint of microscopic-mesoscopic-macroscopic dynamics that we are taking  in this paper,
thermodynamics has to be also seen as a part of  statistical mechanics. Thermodynamics traditionally divides into
classical equilibrium  and  nonequilibrium. In the following four paragraphs we
 briefly recall the main tenets of the four parts of  statistical mechanics.

Equilibrium statistical mechanics is an investigation of the MaxEnt passage from the level of classical (or quantum) mechanics of $\sim 10^{23}$
particles to the equilibrium level with the volume $V$, number of moles $N$  and energy $E$ playing the role of state variables
(see Section \ref{mp} with
the upper thermodynamic relation (\ref{Geq11})). The investigation includes its thermodynamic limit $N\rightarrow \infty,\,\,\,
V\rightarrow \infty,\,\,\, N/V=const.$ in \cite{Ruelle} as well as its various geometrical deformations (e.g. in \cite{Green}) and approximations
\cite{Sethna}.

Nonequilibrium statistical mechanics is a collection of various investigations, both in deterministic and stochastic settings, of the time
evolution taking place on mesoscopic levels \cite{Maes}. The collection has  no clear organizing  principle.

The classical equilibrium thermodynamics \cite{*Gibbs}, \cite{Callen} is a very clearly formulated theory. However, the postulates on which
it stands are disconnected from the more microscopic viewpoints and its domain of applicability is limited to macroscopic systems at equilibrium,
i.e. to macroscopic systems that
are particulary prepared by leaving them sufficiently long time without external influences.

Nonequilibrium thermodynamics is a theory recalled  in Section \ref{fmt}. The difference between nonequilibrium statistical mechanics and
nonequilibrium thermodynamics is only in state variables. If the state variables can be physically interpreted as distribution functions
then it is customary to consider the investigation as a part of statistical mechanics.
For example, if the state variable is the  one particle distribution function then
investigations of its time evolution belong traditionally to kinetic theory that is considered to be a part of the
nonequilibrium statistical mechanics. If however  the state variables are the
first five moments of the one particle distribution function (i.e. the hydrodynamic fields)
then investigations of their time evolution belong traditionally to the nonequilibrium thermodynamics.

\textit{Following  Section \ref{structure}, we suggest to regard statistical mechanics (that we call multiscale thermodynamics) as an oriented
 graph in which vertices are levels and links are reductions. The links point from upper to  lower levels.
 Depending on the features
 of the time evolution that are being compared in the reductions, the links are of three types: state-space, rate, and boundary. Each link
 is then  of two types: time-evolution, and MaxEnt. Altogether, every upper level is connected with a lower level by tree pairs of arrows, all
 pointing  to the lower level}.

 State-space reductions are described in Section \ref{structure}. The essence of the reduction is a recognition of a pattern  in the upper
 phase portrait. The recognized pattern is then identified with the lower phase portrait. We recall that an upper  phase portrait is a collection of
 trajectories generated by an ensamble  of upper vector fields and passing through all points in the upper state space. Similarly, the lower
 phase portrait is a collection of trajectories generated by an ensamble  of lower vector fields and passing through all points
 in the lower state space.

 Rate reductions are described in Section \ref{rt}. They are the same as the state-space reductions
 except that the pattern is searched in rate phase
 portraits rather than in phase portraits. The rate phase portrait is a collection   of rate trajectories in the space of vector
 fields on the state space.
 The rate trajectories are solutions of the time evolution lifted from the state space to the space of vector fields on the state space.

 Boundary reductions are the same type of reductions as the two previous ones  except that the focus is put on the behavior in the region in which
 bulk meets the boundary of  the system under investigation.  In this paper we have not discussed them but this part of statistical mechanics
 is obviously very important and will be pursued in the future.

All three types of reductions  are made by following  reducing time evolutions to their  conclusions. All reducing time evolutions are generated
by a potential called an entropy. Two  different reductions  are  generated (at least in general) by two different entropies.
The entropy increases during the reducing time evolution. The states at which the entropy reaches its maximum
(subjected to constraints determined by the target lower level) are thus the states
representing the lower state variables in the upper state space.  These states can be thus reached either by following the reducing
time evolution to its conclusion or simply by maximizing the entropy subjected to appropriate constraints.
The former way is the time-evolution reduction (see Section \ref{te})
the latter is the MaxEnt reduction (see Section \ref{mp}).

Two vertices in the multiscale-thermodynamics graphs
have a special status. One is with no incoming arrows and the other with no outcoming  arrows. The former is the  most
microscopic level on which macroscopic systems are seen as composed of $\sim 10^{23}$ particles. The latter is the level of the classical
equilibrium thermodynamics on which states are characterized by the volume, he number of moles, and the energy.

The reducing entropies on the
most microscopic level  are potentials, called Casimirs,   that
remain unchanged during the upper time evolution. There are infinitely many  Casimirs.
In order to single out one among them we need to identify
a nucleus of dissipation (e.g collisions in dilute gases). The nucleus increases during the time evolution and the resulting dissipative
time evolution gives rise to  a reduction representred by an outcoming arrow. The reduction is generated by a potential  which is then the entropy
singled out among the infinitely many Casimirs.

The reduced entropies appearing on the level of the classical
equilibrium thermodynamics due to  incoming arrows manifest themselves in the separation  of the total energy into a macroscopic
mechanical energy and heat. The former  can be directly and completely converted into macroscopic mechanical work, the latter only
incompletely in  Carnot's machines.

Seeing the conventional viewpoint of statistical mechanics in the context of the graph of multiscale-thermodynamics, we note that the conventional
viewpoint
concentrates on some particular links (in particular those leaving the microscopic level) while the  graph viewpoint
puts the same attention on all links in the graph. We may  anticipate that
a general theory of the graph (in particular an identification of common features of all the links) can  contribute to a deeper understanding
of the links leaving the microscopic level (which are  obviously  of particular importance for understanding  emerging phenomena).
For instance, it has been suggested in Section \ref{Gibbsevol} that the Grad-Villani dissipation enhancement that seems to be present
in all links may be such a contribution.

\section{Kinematics-preserving Hierarchies}\label{hierarchy}

This section illustrates the use of  multiscale thermodynamics in advancing specific problems.
One of the oldest and probably still the most frequently used strategy to make    reductions   is to begin by casting the upper level
governing equations
  into a hierarchy of equations. The hierarchy reformulation is chosen in such a way that the lower part of the hierarchy  is  already the system of
reduced governing equations   that we look for except that  the equations still remain  coupled to the rest of the hierarchy.
Our objective in this section is to place the hierarchy reductions into the larger context of the multiscale thermodynamics and to
show some of the  implications.

We present the
mathematical formulation first for the case when the upper state variable $x\in M^{\uparrow}$ is the N-particle distribution function $f(1,...,N)$
(we recall that
we use the shorthand notation $1=(\rr_1,\vv_1),...,N=(\rr_N,\vv_N)$). Our objective is to lift the time evolution of $f$ to
space  $M^{\uparrow}$ with the state variables $\mathcal{Z}=(\mathbb{Z},f(1,...,N))\in M^{\uparrow}$, where
\begin{eqnarray}\label{BB1}
\mathbb{Z}=(Z_1,...,Z_n)&=&\left(\int d1...\int dN f(1,...,N) z_1(1,...N),...,\right. \nonumber \\
&&\left. \int d1...\int dN f(1,...,N) z_n(1,...N)\right)
\end{eqnarray}
and $z(1,...,N)=(z_1(1,...,N),...,z_n(1,...,N))$ is a fixed set of $n$ functions.
We recall that reduction is a pattern recognition in the upper phase portrait. We assume that from some previous considerations we already
have a reason to anticipate that   $\mathbb{Z}$ will play an important role in expressing the pattern
(see examples in
Sections \ref{3.4} and \ref{3.2}).

As for the time evolution of $f$ on the level $\mathcal{L}$,
we restrict ourselves to the time evolution equations in the form (\ref{Ham1}). In other words, we consider
the time evolution equations  in the form
\begin{equation}\label{BB2}
\dot{A}=\{A,E\}^{\uparrow}\,\,\,\, \forall A
\end{equation}
where $\{A,B\}^{\uparrow}$ is a Poisson bracket. We thus consider in this section only Hamiltonian dynamics. However,
the contact geometry setting that is discussed in Section \ref{CS}, a slightly modified
Eq.(\ref{BB2})  (see Eq.(7.7) in Ref.\cite{PKGbook}) represents also GENERIC dynamics.
Kinematics-preserving hierarchy formulation of GENERIC dynamics will be explored in a future paper.

Having the time evolution equation (\ref{Ham1}) and the mapping (\ref{BB1}), the first equation
in the \textit{standard hierarchy reformulation} of (\ref{Ham1}) is obtained by multiplying (\ref{Ham1}) by
$z_1(1,...,N)$ and integrating over $\int d1...\int dN$. The second equation is obtained in the same way but with $z_2(1,...,N)$ replacing
$z_1(1,...,N)$. Continuing this process we obtain the standard hierarchy consisting of  $n+1$ time evolution equations; $n$ equations governing the
time evolution of $\mathbb{Z}$ that are coupled to the $(n+1)th$  equation (\ref{Geq}) governing the time evolution of $f$. The next step
in the reduction is the  "closure of the hierarchy" consisting of
 expressing  $f$ in terms of $\mathbb{Z}$. The final reduced dynamics
in $M^{\downarrow}$ consists of $n$ equations governing the time evolution of $\mathbb{Z}$. In the unclosed form the hierarchical reformulation
represents in fact a coupled dynamics of  the upper and the lower levels.
By choosing appropriately the functions $z(1,...,N)=(z_1(1,...,N),...,z_n(1,...,N))$
(see e.g. illustrations in Sections \ref{3.4} and \ref{3.2}), the resulting hierarchy can be made to involve only $\mathbb{Z}$ and not $f$.
The prize to pay for such elimination of the overall state variable $f$
is that $n = \infty$, i.e. the hierarchy is infinite.  The closure in such  case consists of   replacing
  an infinite hierarchy with a finite  hierarchy (see Sections \ref{3.4} and \ref{3.2} below).

In this paper we  take another path. We shall make  an alternative  hierarchy reformulation of (\ref{Ham1}).
We use the fact that  the vector field (\ref{Ham1}) (i.e. the right hand side of (\ref{Ham1})) is composed of two structural elements and  cast
into the hierarchy form only one of them. The vector field (\ref{Ham1}) is
a force (gradient of a potential $E$) transformed into a vector by kinematics
expressed mathematically in the bivector $L^{\uparrow}$ (or equivalently in the Poisson bracket $\{,\}^{\uparrow}$).
In the hierarchy reformulation we concentrate only on the
kinematics, i.e. only on the Poisson bracket $\{,\}^{\uparrow}$. We reformulate it into a  hierarchy    form that retains  the Poisson structure.
The energy $E$ remains in the reformulation undetermined. Since it is the energy where the individual nature of the macroscopic systems is
expressed, the particular physics of the system under investigation does not enter the mathematical modeling  before starting 
the hierarchy reformulation (as it is done in the standard approach)  but after the kinematics-preserving hierarchy reformulation has been made. 
The specification of the energy becomes thus an extra tool in   reductions.

In order to obtain  such \textit{kinematics-preserving reformulation} of the Liouville equation (\ref{Geq}), we proceed as follows.
 The functions $A$ and $B$ in the
Poisson bracket $\{A,B\}$  are assumed to
depend on $f$  both directly and through their dependence on $\mathbb{Z}$ given in (\ref{BB1}). This means
that $A_{f(1,...,N)}$ turns into $z_{\alpha}(1,...,N)A_{Z_{\alpha}} +A_f $, where $\alpha=1,...,n$  (the summation convention
over repeated indices is used). With these expressions
for $A_f$ and $B_f$, the Poisson bracket (\ref{Gkin}) becomes
\begin{eqnarray}\label{Z1}
\{A,B\}&=&\int d1...\int dN f \left[ \frac{\partial z_{\alpha}}{\partial r_{\gamma i}}\frac{\partial z_{\beta}}{\partial v_{\gamma i}}
\left(A_{Z_{\alpha}}B_{Z_{\beta}}-B_{Z_{\alpha}}A_{Z_{\beta}}\right)\right.\nonumber \\
 && \left. + \frac{\partial z_{\alpha}}{\partial r_{\gamma i}}\left(A_{Z_{\alpha}}\frac{\partial B_f}{\partial v_{\gamma i}}-
B_{Z_{\alpha}}\frac{\partial A_f}{\partial v_{\gamma i}}\right)\right]\nonumber \\
&& \left. + \frac{\partial z_{\alpha}}{\partial v_{\gamma i}}\left(B_{Z_{\alpha}}\frac{\partial A_f}{\partial r_{\gamma i}}-
A_{Z_{\alpha}}\frac{\partial B_f}{\partial r_{\gamma i}}\right)\right.\nonumber \\
&&\left. +\left(\frac{\partial A_f}{\partial r_{\gamma i}}\frac{\partial B_f}{\partial v_{\gamma i}}-
\frac{\partial B_f}{\partial r_{\gamma i}}\frac{\partial A_f}{\partial v_{\gamma i}}\right)\right]
\end{eqnarray}
The time evolution equations (\ref{BB2}) with $\{A,B\}^{\uparrow}$ given in (\ref{Z1}) take the form
\begin{eqnarray}\label{Z2}
\dot{Z}_{\alpha}&=& \int d1...\int dN f\left[\left(\frac{\partial z_{\alpha}}{\partial r_{\gamma i}}\frac{\partial z_{\beta}}{\partial v_{\gamma i}}-
\frac{\partial z_{\beta}}{\partial r_{\gamma i}}\frac{\partial z_{\alpha}}{\partial v_{\gamma i}}\right)E_{Z_{\beta}}\right.\nonumber \\
&&\left.+\left(\frac{\partial z_{\alpha}}{\partial r_{\gamma i}}\frac{\partial E_f}{\partial v_{\gamma i}}-
\frac{\partial z_{\alpha}}{\partial v_{\gamma i}}\frac{\partial E_f}{\partial r_{\gamma i}}\right)\right]\nonumber \\
\frac{\partial f}{\partial t}&=& -\frac{\partial}{\partial r_{\gamma i}}\left(f \frac{\partial z_{\alpha}}{\partial v_{\gamma i}}\right)E_{Z_{\alpha}}+
\frac{\partial}{\partial v_{\gamma i}}\left(f \frac{\partial z_{\alpha}}{\partial r_{\gamma i}}\right)E_{Z_{\alpha}}\nonumber \\
&& -\frac{\partial}{\partial r_{\gamma i}}\left(f \frac{\partial E_f}{\partial v_{\gamma i}}\right)+
\frac{\partial}{\partial v_{\gamma i}}\left(f \frac{\partial E_f}{\partial r_{\gamma i}}\right)
\end{eqnarray}

Summing up, we have cast (\ref{Geq}) into the form (\ref{Z2}). Both equations (\ref{Geq}) and  (\ref{Z2}) are Hamilton's equations,
both are particular realizations of
(\ref{BB2}). The reason why we have passed  from (\ref{Geq}) to (\ref{Z2}) is that
the latter equation is more suitable for starting the reduction process.
We assume we know from some other considerations (for instance from experimental observations)
that the pattern that represents the lower level  in the phase portrait of (\ref{Geq}) can be expressed in terms of $\mathbb{Z}$.
If this is the case then clearly the reformulation (\ref{Z2}) of (\ref{Geq}) is more suitable for investigating the reduction. Both
(\ref{Z2}) and  (\ref{Geq}) share the same kinematics but the energies in them
remain so far completely unrelated and at this point undetermined.
Their determination  is a part of the continuation of
the pattern recognition process in the phase portrait of (\ref{Geq}) that has to enter into an actual analysis of solutions to (\ref{Geq}).

In this paper we  make
only a few comments about physical aspects of the hierarchy
(\ref{Z2}). Let $\mathbb{Z}$ be the state variables used on the lower level. The hierarchy (\ref{Z2})  thus governs the time
evolution on the lower level.  However, the  time evolution governed by (\ref{Z2}) is still coupled to the time evolution of $f$.
We  can physically interpret $f$  as a state variable characterizing  overall features of
the  phase portrait of (\ref{Geq}) that are not expressed in the lower state variables $\mathbb{Z}$. An example of considerations, based on
physical assumptions and approximations,  that lead to expressing $f$ in terms of $\mathbb{Z})$  is presented in the next section.

Still another insight into the physics expressed in  the hierarchy (\ref{Z2}) is revealed in its following reformulation. We note that the last
equation in (\ref{Z2}) (i.e. the equation governing the time evolution of $f$) can be seen as the Liouville (i.e. continuity)
equation corresponding to 6N
ordinary differential equations governing the time evolution of $(1,...,N)$. We can thus reformulate (\ref{Z2}) into a system $(n+6N)$
ordinary differential equations
\begin{eqnarray}\label{Z3}
\dot{Z}_{\alpha}&=& \int d1...\int dN f\left[\left(\frac{\partial z_{\alpha}}{\partial r_{\gamma i}}\frac{\partial z_{\beta}}{\partial v_{\gamma i}}-
\frac{\partial z_{\beta}}{\partial r_{\gamma i}}\frac{\partial z_{\alpha}}{\partial v_{\gamma i}}\right)E_{Z_{\beta}}\right.\nonumber \\
&&\left.+\left(\frac{\partial z_{\alpha}}{\partial r_{\gamma i}}\frac{\partial E_f}{\partial v_{\gamma i}}-
\frac{\partial z_{\alpha}}{\partial v_{\gamma i}}\frac{\partial E_f}{\partial r_{\gamma i}}\right)\right]\nonumber \\
\dot{r}_{\gamma i}&=&\frac{\partial z_{\alpha}}{\partial v_{\gamma i}}E_{Z_{\alpha}}+\frac{\partial E_f}{\partial v_{\gamma i}}\nonumber \\
\dot{v}_{\gamma i}&=&-\frac{\partial z_{\alpha}}{\partial r_{\gamma i}}E_{Z_{\alpha}}-\frac{\partial E_f}{\partial r_{\gamma i}}
\end{eqnarray}
that is accompanied with
\begin{equation}\label{Z4}
f(1,...,N,t)=f_0(\mathbb{T}_{-t}(1,...,N))
\end{equation}
where $\mathbb{T}_t$  is the trajectory of $(1,...,N)$ and $f_0(1,...,N)$ is an initial distribution function. In this formulation we
see clearly  the role of $f$. It is indeed a state variable expressing overall features of the dynamics that are expressed
neither in $(1,...,N)$ nor in $\mathbb{Z}$.

Finally, we emphasize that all the reformulations that we have made above in this section do not involve  any approximation. Both (\ref{Geq}) and
(\ref{Z2}) share the same kinematics (but in different representations) and the energies in both equations remain undetermined.
We can see (\ref{Z2}) as a combination of the microscopic level represented by (\ref{Geq}) and the mesoscopic level on which $\mathbb{Z}$
serve as state variables. In the next section we makes initial steps in the pattern recognition process leading to the closure of (\ref{Z2}) (i.e.
to expressing $f$ in terms of $\mathbb{Z})$.

\subsection{Dissipation, Closure}\label{DC}

In order to recognize a pattern in the phase portrait corresponding to (\ref{Geq}) or to its reformulation (\ref{Z2}), the phase portrait
has to be first created (i.e. solutions to (\ref{Geq}) or to (\ref{Z2})  have to be found).
This can be done only if we  specify the energy  $E^{\uparrow}(x)$
entering (\ref{Geq}) or (\ref{Z2})
and thus commit ourselves to specific macroscopic systems. An example of such  analysis is in \cite{Villani} for the Boltzmann equation.
We shall not follow this path. Instead, we limit ourselves  to some observations of a  qualitative nature that combine
physical and mathematical arguments.

First,  we emphasize that the choice (\ref{BB1}) of  the lower state variables $\mathbb{Z}$ is already a part of the pattern recognition process.
We anticipate that
the pattern that we search  in the phase portrait corresponding to (\ref{Geq})  can  be expressed in terms of $\mathbb{Z}$. For example,  we
may recall the considerations leading to the choice of the hydrodynamic fields. Since the total mass, momentum and energy are conserved,
the local mass, momentum and energy change in slower pace than other mesoscopic state variables.

Having chosen  (\ref{BB1}), we follow the previous section and arrive at the hierarchy (\ref{Z2}) (or (\ref{Z3})) combining the upper and lower levels.
Now we make a very obvious but important observation.
We can look at the hierarchy in two ways: "bottom up" and "top down". In (\ref{Z2}) the bottom part is the first
equation and the top part is the second equation. In infinite hierarchies the second equation is replaced by an infinite hierarchy of equations
governing the time evolution of $(Z_{n+1}, Z_{n+2},...)$.

The standard view is  bottom up. It is the reduced dynamics,  which is an appropriately closed bottom part of the hierarchy,
that is typically  the reason why the reductions are made.
We  look  at the bottom part of the hierarchy and try to close it. In the hierarchy (\ref{Z2}) this means that
we  look at the first equation  and  try  to express $f$ appearing in it in terms of $\mathbb{Z}$
(or in infinite hierarchies in terms of $(Z_{n+1}, Z_{n+2},...))$.
The closure can  be argued by putting various requirements on the closed system of equations. In our analysis the natural
requirement is that the closed system of equations  remains to be a system of Hamilton's equations. In the case of dissipative dynamics we
may  require that an appropriate entropy (a real valued function of $\mathbb{Z}$) will not decrease during the time evolution governed by the
closed system of equations. This latter requirement first appeared in \cite{Coleman} and was later developed in \cite{LiuMuller}.

But the closure can also be argued from top down. As we have already noticed in Section \ref{trans}, this viewpoint of the closure is in fact
present in the Chapman-Enskog method and it has also been compared with the bottom up viewpoint in \cite{Dreyer} (where however  only
the static MaxEnt version of the top part of the hierarchy is considered). In this paper we  continue to explore the
top down view of the closure.

The top part of the hierarchy (\ref{Z2}) is its    second equation  in which \\$z_1(1,...,N),...,z_n(1,...,N)$ are seen as quantities
representing external influences in the time evolution of $f$. If (\ref{Z2}) is cast into  the form (\ref{Z3}) then  the
top part  consists of  the last two equations in (\ref{Z3}) together with (\ref{Z4}). The quantities  $\mathbb{Z}$ appear in (\ref{Z3}) indeed
as extra velocities and extra forces in the Hamilton time evolution of $(1,...,N)$. In order to close the hierarchy (\ref{Z2})
we have to find the phase portrait corresponding
to its top part  and then recognize in it  a pattern parametrized by $\mathbb{Z}$. In other words, we have  to express
$f$ in terms of $\mathbb{Z}$ (we denote it $\hat{f}(\mathbb{Z};1,...,N)$) by analyzing solutions of the top part of the hierarchy.
By inserting        $\hat{f}(\mathbb{Z};1,...,N)$  into the lower part of the hierarchy (i.e.
into the first equation in (\ref{Z2})), we  arrive at  the lower dynamics.

Such  investigation  cannot be done without making  commitment to a specific physical system and without a type of
analysis displayed for example in \cite{Villani}.  In the rest of this section we make only a very
qualitative analysis that transforms the bottom part of (\ref{Z2}) with  the Onsager time evolution equation.

We begin by noting that $S(f)=\int d1...\int dN \eta(f)$, where $\eta:\mathbb{R}\rightarrow \mathbb{R}$ is a sufficiently regular function,
 is a Casimir of the Poisson bracket (\ref{Z1}). We can therefore put (\ref{Z3}) into the form
\begin{eqnarray}\label{Z13}
\dot{Z}_{\alpha}&=& \int d1...\int dN f\left[\left(\frac{\partial z_{\alpha}}{\partial r_{\gamma i}}\frac{\partial z_{\beta}}{\partial v_{\gamma i}}-
\frac{\partial z_{\beta}}{\partial r_{\gamma i}}\frac{\partial z_{\alpha}}{\partial v_{\gamma i}}\right)\Phi_{Z_{\beta}}\right.\nonumber \\
&&\left.+\left(\frac{\partial z_{\alpha}}{\partial r_{\gamma i}}\frac{\partial \Phi_f}{\partial v_{\gamma i}}-
\frac{\partial z_{\alpha}}{\partial v_{\gamma i}}\frac{\partial \Phi_f}{\partial r_{\gamma i}}\right)\right]\nonumber \\
\frac{\partial f}{\partial t}&=& -\frac{\partial}{\partial r_{\gamma i}}\left(f \frac{\partial z_{\alpha}}{\partial v_{\gamma i}}\right)
\Phi_{Z_{\alpha}}+
\frac{\partial}{\partial v_{\gamma i}}\left(f \frac{\partial z_{\alpha}}{\partial r_{\gamma i}}\right)\Phi_{Z_{\alpha}}\nonumber \\
&& -\frac{\partial}{\partial r_{\gamma i}}\left(f \frac{\partial \Phi_f}{\partial v_{\gamma i}}\right)+
\frac{\partial}{\partial v_{\gamma i}}\left(f \frac{\partial \Phi_f}{\partial r_{\gamma i}}\right)
\end{eqnarray}
where $\Phi(\mathbb{Z},f)=-S(f)+\frac{1}{T}E(\mathbb{Z},f)$ and the  constant temperature $T$ is absorbed in rescaling the time. In the next step
we introduce to the
top equation in the hierarchy (i.e. to the second equation in (\ref{Z13})) a dissipation. From physical considerations we
anticipate that  the dominant dissipation is the Fokker-Planck type diffusion in momenta. If we restrict ourselves to the linear dissipation
(i.e. if we restrict ourselves to states that are  not far from  equilibrium states), the top part of the hierarchy (\ref{Z13}) becomes
\begin{eqnarray}\label{Z14}
\frac{\partial f}{\partial t}&=& -\frac{\partial}{\partial r_{\gamma i}}\left(f \frac{\partial z_{\alpha}}{\partial v_{\gamma i}}\right)
\Phi_{Z_{\alpha}}+
\frac{\partial}{\partial v_{\gamma i}}\left(f \frac{\partial z_{\alpha}}{\partial r_{\gamma i}}\right)\Phi_{Z_{\alpha}}\nonumber \\
&& -\frac{\partial}{\partial r_{\gamma i}}\left(f \frac{\partial \Phi_f}{\partial v_{\gamma i}}\right)+
\frac{\partial}{\partial v_{\gamma i}}\left(f \frac{\partial \Phi_f}{\partial r_{\gamma i}}\right)\nonumber \\
&&-\frac{\partial}{\partial v_{\gamma i}}\left(f\lambda_{\gamma \delta i j}\frac{\partial \Phi_f}{\partial v_{\delta j}}\right)
\end{eqnarray}
where $\lambda$ is a symmetric matrix guaranteeing
$\int d1...\int dN  f\frac{\partial \Phi_f}{\partial v_{\gamma i}}\lambda_{\gamma \delta i j}\frac{\partial \Phi_f}{\partial v_{\delta j}}\geq 0$

To continue,  we use  physical considerations to identify dominant terms on the right hand side of (\ref{Z14}).
We recall that this type of arguments
is also the point of departure of the Chapman-Enskog passage from the Boltzmann kinetic equations to fluid mechanics (see Section \ref{Boltzmann}).
The anticipated dominance
of variations in momenta (which  can be physically interpret  as an anticipation of the occurrence  of turbulence on the micro scale),
that led us already to the introduction of the
dissipative term in (\ref{Z14}),  leads us to regard the second term on the right hand side of (\ref{Z14}) as  dominant.
We thus assume that the main part of
(\ref{Z14}) is the time evolution equation
\begin{equation}\label{Z15}
\frac{\partial f}{\partial t}=\frac{\partial}{\partial v_{\gamma i}}\left(f \frac{\partial z_{\alpha}}{\partial r_{\gamma i}}\right)\Phi_{Z_{\alpha}}
-\frac{\partial}{\partial v_{\gamma i}}\left(f\lambda_{\gamma \delta i j}\frac{\partial \Phi_f}{\partial v_{\delta j}}\right)
\end{equation}
We have omitted the term
$\frac{\partial}{\partial r_{\gamma i}}\left(f \frac{\partial \Phi_f}{\partial v_{\gamma i}}\right)$ that we assume to be smaller than the remaining
two terms. Next, we assume that the distribution function $f$ evolves faster than $\mathbb{Z}$ and that we are already
in the stage of the time evolution  in which $f$  reached  the stationary state.

We need now to solve
\begin{equation}\label{Z15}
0=\frac{\partial}{\partial v_{\gamma i}}\left(f \frac{\partial z_{\alpha}}{\partial r_{\gamma i}}\right)\Phi_{Z_{\alpha}}
-\frac{\partial}{\partial v_{\gamma i}}\left(f\lambda_{\gamma \delta i j}\frac{\partial \Phi_f}{\partial v_{\delta j}}\right)
\end{equation}
In order to avoid complications with the degeneracy of the matrix $\lambda$  (that is needed to
satisfy the energy conservation ), we limit ourselves
in this paper to isothermal systems. The thermodynamic potential $\Phi$ is thus the Helmholtz free energy and the matrix $\lambda$
 is a positive definite matrix which can be inverted. We can thus easily solve (\ref{Z15}). If we insert the solution
into the first equation in (\ref{Z13}) (in which we omit the term
$\frac{\partial}{\partial r_{\gamma i}}\left(f \frac{\partial z_{\alpha}}{\partial v_{\gamma i}}\right)$ that we assume to be small relative
to the other terms in (\ref{Z13})), we obtain
\begin{equation}\label{O1}
\dot{Z}_{\alpha}=L_{\alpha \beta}\Phi_{Z_{\beta}}-\Lambda_{\alpha \beta}\Phi_{Z_{\beta}}
\end{equation}
where
\begin{eqnarray}\label{O2}
L_{\alpha \beta}&=&\int d1...\int dN f\left(\frac{\partial z_{\alpha}}{\partial r_{\gamma i}}\frac{\partial z_{\beta}}{\partial v_{\gamma i}}-
\frac{\partial z_{\beta}}{\partial r_{\gamma i}}\frac{\partial z_{\alpha}}{\partial v_{\gamma i}}\right)\\\label{O3}
\Lambda_{\alpha \beta}&=&\int d1...\int dN f \frac{\partial z_{\alpha}}{\partial r_{\gamma i}}\lambda^{-1}_{\gamma \delta, i j}
\frac{\partial z_{\beta}}{\partial r_{\delta j}}
\end{eqnarray}

Equation (\ref{O1}) governing  the time evolution of $\mathbb{Z}$ still needs to be closed. The distribution functions $f$ still appears in
the matrices $L$ and $\Lambda$ in (\ref{O2}) and  (\ref{O3}). We assume now that we have an independent information about the overall  state
of the system under investigation and thus about $f$. For instance, if the system under investigation is close to equilibrium, we can
replace $f$ in (\ref{O2}), (\ref{O3}) by the Gibbs equilibrium distribution function.

Considering $f$ in  (\ref{O2}) and  (\ref{O3})  known, Eq.(\ref{O1}) is a closed equation governing the time evolution of the
mesoscopic state variables $\mathbb{Z}$. Equation (\ref{O1}) is the Onsager equation Ref.\cite{Onsager}. The matrix $\Lambda$ is symmetric,
positive definite,  and it does not change its sign if the momenta $(\vv_1,...,\vv_N)$ change their signs. The matrix $L$ is skewsymmetric and it
changes its sign if $(\vv_1,...,\vv_N)\rightarrow (-\vv_1,...,-\vv_N)$. Equation (\ref{O1}) is also a GENERIC equation since (\ref{O1}) is
a particular realization of (\ref{GEN}). The matrix $L$ is indeed skewsymmetric and the Poisson bracket $\{A,B\}=$
$A_{Z_{\alpha }}L_{\alpha \beta }
B_{Z_{\beta }}$ satisfies the Jacobi identity since $L$ is independent of $\mathbb{Z}$. We note that the
only unspecified
parameter in the formulas (\ref{O2}),(\ref{O3}) for the matrices $L$ and $\Lambda$
is  the matrix $\lambda$ (entering the microscopic time evolution (\ref{Z15})).

Summing up,
the Onsager \cite{Onsager}   result (\ref{O1}) has emerged as a part of the reduced structure on the lower level
with the state variables $\mathbb{Z}$. The reduction is made by
reducing time evolution   that preserves the Hamiltonian structure of the time evolution that takes place on the upper level. We note in
particular that the Onsager symmetry of $\Lambda$ is a direct consequence of the gradient structure on the upper level that guarantees the
existence of the reduction. The skewsymmetry of $L$ and its  sign change in the transformation $(\vv_1,...,\vv_N)\rightarrow
(-\vv_1,...,-\vv_N)$ is a direct consequence of
the Hamiltonian structure of the upper level time evolution. In the absence of dissipation,  Eq.(\ref{O1}) represents GENERIC dynamics.

We can also find the lower rate thermodynamic relation $\Sigma^{\downarrow}(y)$ implied by the reduction discussed above. We note that
with
\begin{equation}\label{Z20}
\Sigma^{\uparrow}(X)=-\frac{1}{2}\int d1...\int dN f X_{\gamma i}\lambda_{\gamma \delta i j}X_{\delta j}
\end{equation}
we can write (\ref{Z15}) as
\begin{equation}\label{Z26}
\Psi^{\uparrow}_X(X,Y)=0
\end{equation}

where
\begin{equation}\label{Z27}
\Psi^{\uparrow}(X,Y)=-\Sigma^{\uparrow}(X)+\int d1...\int dN f Y_{\gamma i}X_{\gamma i}
\end{equation}
with
\begin{equation}\label{Z28}
Y_{\gamma i}=\frac{\partial z_{\alpha}}{\partial r_{\gamma i}}\Phi_{Z_{\alpha}}
\end{equation}

Consequently, the lower rate entropy implied by above reduction is
\begin{equation}\label{Z29}
\Sigma^{\downarrow}(Y)=\frac{1}{2}\int d1...\int dN f \frac{\partial z_{\alpha}}{\partial r_{\gamma i}}\lambda^{-1}_{\gamma \delta, i j}
\frac{\partial z_{\beta}}{\partial r_{\delta j}}
\end{equation}
We note that Eq.(\ref{O1}) implies
\begin{equation}\label{Z1000}
\dot{\Phi}=\Phi_{Z_{\alpha}}\dot{Z}_{\alpha}=\frac{1}{2}\Sigma^{\downarrow}
\end{equation}
which relates the  lower rate entropy to the lower entropy production.

\subsection{Illustrations}\label{illustrations}

The Liouville equation (\ref{Geq}) governing the
Hamiltonian time evolution of N-particle distribution function, $N\sim 10^{23}$,  was the first  equation that was cast into the hierarchy form.
The hierarchy reformulation of the Liouville equation (\ref{Geq}) is called BBGKY hierarchy \cite{BBGKY1}, \cite{BBGKY2}.
Another time evolution equation that gave rise to a famous hierarchy (Grad hierarchy \cite{Grad}) is the Boltzmann kinetic
equation. Below, we shall cast into hierarchy also the Euler hydrodynamic equation. The kinematics-preserving hierarchies for all three equations
illustrate the general analysis presented above.  In all three hierarchies we limit ourselves in this paper
only to the Hamiltonian part of the time evolution.

\subsubsection{BBGKY kinematics-preserving hierarchy}\label{3.4}

The state variable on the upper level is the N-particle distribution function $f_N(1,...,N)$. The lower level state variables are $1,...,N-1$
distribution functions obtained from $f_N$ by integrating over the $N-1,...,2$ coordinates respectively. All distribution functions
are symmetric with respect to to the relabeling the particles.
The kinematics of $f_N$ is expressed in the Poisson bracket
(\ref{Gkin}).

The infinite kinematics-preserving hierarchy in this setting has been developed in \cite{GrmelaBBGKY} for $1,...,N,...$ distribution functions.
Here we present a kinematics-preserving hierarchy reformulation of the Liouville equation governing the time evolution of $f_2(1,2)$.
To simplify the notation, we use  $f$ to denote $f_2$ and $g$ to denote $f_1$. The one particle distribution function $g$ is related to $f$ by
\begin{equation}\label{fg}
g(\rr,\vv)=\int d1\int d2 f(1,2) \left[\delta(\rr-\rr_1)\delta(\vv-\vv_1)+\delta(\rr-\rr_2)\delta(\vv-\vv_2)\right]
\end{equation}

In order to find the kinematics of $(f,g)$ we take the Poisson bracket (\ref{Gkin}) for $N=2$  with the functions $A$ and $B$ that depend on $f$
directly and also indirectly through their dependence on $g$ (that depends on $f$ - see (\ref{fg})).
By  replacing  $A_f$ in  (\ref{Gkin}) with
$A_{f(1,2)}= A_{g(1)}+A_{g(2)}+A_{f(1,2)}$ and similarly $B_{f(1,2)}$, the Poisson bracket (\ref{Gkin}) becomes the Poisson bracket
\begin{eqnarray}\label{BBB}
\{A,B\}&=&\{A,B\}^{(N=1)}\nonumber \\
&&+\int d1 \int d2\left[f\left(\frac{\partial A_{f}}{\partial r_{1i}}\frac{\partial B_{g(1)}}{\partial v_{1i}}-
\frac{\partial B_{f}}{\partial r_{1i}}\frac{\partial A_{g(1)}}{\partial v_{1i}}\right)\right.\nonumber \\
&&+\left.f\left(\frac{\partial A_{f}}{\partial r_{2i}}\frac{\partial B_{g(2)}}{\partial v_{2i}}-
\frac{\partial B_{f}}{\partial r_{2i}}\frac{\partial A_{g(2)}}{\partial v_{2i}} \right)\right.\nonumber \\
&&+\left.f\left(\frac{\partial A_{g(1)}}{\partial r_{1i}}\frac{\partial B_{f}}{\partial v_{1i}}-
\frac{\partial B_{g(1)}}{\partial r_{1i}}\frac{\partial A_{f}}{\partial v_{1i}}\right)\right.\nonumber \\
&&+\left.f\left(\frac{\partial A_{g(2)}}{\partial r_{2i}}\frac{\partial B_{f}}{\partial v_{2i}}-
\frac{\partial B_{g(2)}}{\partial r_{2i}}\frac{\partial A_{f}}{\partial v_{2i}}\right)\right]\nonumber \\
&&+\{A,B\}^{(N=2)}
\end{eqnarray}
where $\{A,B\}^{(N=1)}$ and $\{A,B\}^{(N=2)}$ are the Poisson brackets (\ref{Gkin}) for $N=1$ and $N=2$ respectively. There is an important difference
between the Poisson bracket (\ref{BBB}) and the bracket (\ref{Z1}). In (\ref{BBB}) the  bracket is a sum of three brackets, one involving only
 the lower  state variable $g$, the other involve both the upper $f$ and the lower $g$ state variables, and the third involves only
 the upper state variable $f$, The bracket  (\ref{Z1}) is a sum of two terms, one involving both the upper and the lower state variables and the other
 only the upper state variable.

The time evolution equation (\ref{BB2}) corresponding to the Poisson bracket (\ref{BBB}) becomes
\begin{eqnarray}\label{eqsbb}
\frac{\partial g}{\partial t}&=&-\frac{\partial}{\partial r_i}\left(g\frac{\partial E_g}{\partial v_i}\right)+\frac{\partial}{\partial v_i}\left
(g\frac{\partial E_g}{\partial r_i}\right)\nonumber \\
&&+2\int d\rr_2\int d\vv_2\left[-\frac{\partial}{\partial r_i}\left(f\frac{\partial E_f}{\partial v_i}\right)
+\frac{\partial}{\partial v_i}\left(f\frac{\partial E_f}{\partial r_i}\right)\right]\nonumber \\
\frac{\partial f}{\partial t}&=&\sum_{\alpha=1}^2\left[\frac{\partial}{\partial r_{\alpha i}}\left(f\frac{\partial E_g}{\partial v_{\alpha i}}\right)
+\frac{\partial}{\partial v_{\alpha i}}\left(f\frac{\partial E_g}{\partial r_{\alpha i}}\right)\right]\nonumber \\
&&+\sum_{\alpha=1}^{2}\left[-\frac{\partial}{\partial r_{\alpha i}}\left(f\frac{\partial E_f}{\partial v_{\alpha i}}\right)
+\frac{\partial}{\partial v_{\alpha i}}\left(f\frac{\partial E_f}{\partial r_{\alpha i}}\right)\right]
\end{eqnarray}
The energy $E(f,g)$ remains in these equations undetermined.

Now we discuss qualitative properties of solutions to (\ref{eqsbb}). We note that if the energy $E(f,g)$ is chosen
to be independent of $f$ then (\ref{eqsbb}) turns
into the standard one particle Hamiltonian kinetic equation. This is a consequence of the presence of the bracket $\{A,B\}^{(N=1)}$
(that does not involve $f$) on the right hand side of (\ref{BBB}).
If, on the other hand,  the energy $E(f,g)$ is independent of $g$ then (\ref{eqsbb})
becomes the two particle Hamiltonian kinetic equation. For the energy $E$ that depends on both $g$ and $f$, the time evolution of the
one and the two particle distribution functions are coupled and the two particle distribution function $f$ represents extra details that are
ignored in the one particle kinetic theory.

We also note that the kinematics-preserving  hierarchy (\ref{eqsbb}) is different from the classical BBGKY hierarchy
\begin{eqnarray}\label{Clasbbgky}
\frac{\partial g(\rr,\vv)}{\partial t}&=&-\frac{\partial}{\partial r_i}\left(2\int d2 f(\rr,\vv,\rr_2,\vv_2)\frac{\partial E_{f(\rr,\vv,
\rr_2,\vv_2)}}{\partial v_i}\right)\nonumber \\&& + \frac{\partial}{\partial v_i}\left(2\int d2 f(\rr,\vv,\rr_2,\vv_2)\frac{\partial E_{f(\rr,\vv,
\rr_2,\vv_2)}}{\partial r_i}\right)\nonumber \\
\frac{\partial f(1,2)}{\partial t}&=&\sum_{\alpha=1}^{2}\left[-\frac{\partial}{\partial r_{\alpha i}}
\left(f\frac{\partial E_f}{\partial v_{\alpha i}}\right)+\frac{\partial}{\partial v_{\alpha i}}
\left(f\frac{\partial E_f}{\partial r_{\alpha i}}\right)\right]
\end{eqnarray}
where $E$ is a function of $f$.
The classical BBGKY hierarchy  is obtained from
Eq.(\ref{Geq}) with $N=2$ by simply integrating it over $\int d2$. We recall that the point of departure for obtaining  the kinematics-preserving
hierarchy (\ref{eqsbb}) is the kinematics (\ref{Gkin}) with $N=2$ while the point of departure for the classical hierarchy (\ref{Clasbbgky}) is
the 2-particle Liouville equation (\ref{Geq}).
The original equation (\ref{Geq}) with $N=2$ is
Hamiltonian, its classical hierarchy reformulation (\ref{Clasbbgky}) is not Hamiltonian but its kinematics-preserving hierarchy (\ref{eqsbb})
keeps the  Hamiltonian structure.

\subsubsection{Grad kinematics-preserving hierarchy}\label{3.2}

For the second illustration we turn to the Boltzmann equation governing the time evolution of the one particle
distribution function $f(\rr,\vv)$ and to the time evolution of its Grad's moments
\begin{equation}\label{Grad}
f(\rr,\vv)\mapsto (c^{(0)}(\rr), c^{(1)}_i(\rr),...,c^{(k)}_{i_1,...,i_k}(\rr),...)
\end{equation}
where
\begin{equation}\label{Gradmom}
c^{(k)}_{i_1,...,i_k}(\rr)=\int d\vv v_{i_1}...v_{i_k}f(\rr,\vv)
\end{equation}
The kinematics of $f$ is expressed in the Poisson bracket (\ref{Gkin}) with $N=1$.

The infinite kinematics-preserving hierarchy with an infinite number of Grad's moments (\ref{Gradmom}) as lower state variables
has been worked out in \cite{GrGrad}. Here we present  a Grad kinematics-preserving  5 moment hierarchy   closed by an
equation governing the time evolution of $f$.

We choose (\ref{Grad}) and (\ref{Gradmom}) with
\begin{eqnarray}\label{PH10}
&&f(\rr,\vv)\mapsto (\rho(\rr), \uu(\rr),s(\rr))\nonumber \\
&&\rho(\rr)=\int d\vv f(\rr,\vv);\,\,\uu(\rr)=\int d\vv \vv f(\rr,\vv);\,\,s(\rr)=\int d\vv \eta(f(\rr,\vv))\nonumber \\
\end{eqnarray}
where $\eta(f)$ is a sufficiently regular function $\mathbb{R}\rightarrow \mathbb{R}$. We recall that (see Section \ref{Boltzmann}) that
$\int d\rr\int d\vv \eta(f)$ is a Casimir (see Section \ref{HHam}) of the Poisson bracket (\ref{Gkin}) with $N=1$.
We choose the hydrodynamic
state variables in the energy representation (i.e. the state variables are the fields $(\rho(\rr),\uu(\rr),s(\rr))$ denoting the mass,
momentum, and entropy) rather than in the entropy representation with the state variables
$(\rho(\rr),\uu(\rr),e(\rr)$, where $e(\rr)$ is the energy field. The reason for the choice is explained in Section \ref{fmk}.

From the kinematics of the kinetic theory that is expressed in the Poisson bracket (\ref{Gkin}) with $N=1$ we derive the kinematics-preserving
hierarchy in the same way as in the previous two sections. The functions $A$ and $B$ in  (\ref{Gkin}) with $N=1$  depend on $f$
directly and also through their dependence on the moments (\ref{PH10}). Consequently, we replace $A_f$ and $B_f$ with
$A_{\rho(\rr)}+v_iA_{u_i(\rr)}+\eta_{f(\rr,\vv)}A_{s(\rr)}+A_{f(\rr,\vv)}$ and the same  expression for $B_f$. Straightforward calculations
then lead to the Poisson bracket
\begin{eqnarray}\label{kh}
\{A,B\}^{(kh)}&=&\{A,B\}^{(hyd)}\nonumber \\
&&+\int d\rr\int d\vv \left[f\left(\frac{\partial A_f}{\partial r_i}B_{u_i}-\frac{\partial B_f}{\partial r_i}A_{u_i}\right)\right.
\nonumber \\
&&\left.+f\frac{\partial \eta_f}{\partial v_i}\left(\frac{\partial A_f}{\partial r_i}B_{s}-\frac{\partial B_f}{\partial r_i}A_{s}\right)\right.
\nonumber \\
&&\left.+f\left(\frac{\partial A_{\rho}}{\partial r_i}\frac{\partial B_f}{\partial v_i}-\frac{\partial B_{\rho}}{\partial r_i}
\frac{\partial A_f}{\partial v_i}\right)\right.\nonumber \\
&&\left.+fv_j\left(\frac{\partial A_{u_j}}{\partial r_i}\frac{\partial B_f}{\partial v_i}-\frac{\partial B_{u_j}}{\partial r_i}
\frac{\partial A_f}{\partial v_i}\right)\right.\nonumber \\
&&\left.+f\left(\frac{\partial (A_s\eta_f)}{\partial r_i}\frac{\partial B_f}{\partial v_i}-\frac{\partial (B_s\eta_f)}{\partial r_i}
\frac{\partial A_f}{\partial v_i}\right)\right]\nonumber \\
&&+\{A,B\}^{(N=1)}
\end{eqnarray}
where $\{A,B\}^{(hyd)}$ is the Poisson bracket (\ref{fm3}) expressing kinematics of fluids and $\{A,B\}^{(N=1)}$ is the Poisson bracket
(\ref{Gkin}) with $N=1$. Like the Poisson bracket (\ref{BBB}), but unlike the Poisson bracket (\ref{Z1}), the Poisson bracket (\ref{kh})
 is a sum of three brackets, one depending only on the lower, the other only on the upper,  and  only the third on both the lower and
 the upper state variables. As we have already discussed in Section \ref{fmk},  the hydrodynamic bracket  $\{A,B\}^{(hyd)}$ appears in the
 kinematics-preserving hierarchy bracket only with the hydrodynamic moments (\ref{PH10}).
  With any other choice of the moments (e.g. if the entropy fields $s(\rr)$
 is replaced by the energy field $e(\rr)=\int d\vv \epsilon(\rr,\vv)f(\rr,\vv)$, where $\epsilon(\rr,\vv)$ is a microscopic energy) the
 one particle distribution function $f$ will still remain  in all terms in the Poisson bracket.

The time evolution equations (\ref{Ham1}) with the Poisson bracket (\ref{kh}) and energy
\begin{equation}\label{enfl}
 E(f,\rho,\uu,s)=\int d\rr e(f,\rho,\uu,s;\rr)=\int d\rr \int d\vv \epsilon(f,\rho,\uu,s;\rr,\vv)
\end{equation}
are
\begin{eqnarray}\label{eqs11}
\frac{\partial \rho}{\partial t}&=&-\frac{\partial}{\partial r_i}\left(\rho E_{u_i}+
\int d\vv f\frac{\partial E_f}{\partial v_i}\right)\nonumber \\
\frac{\partial s}{\partial t}&=&-\frac{\partial}{\partial r_i}\left(s E_{u_i}+
\int d\vv \eta\frac{\partial E_f}{\partial v_i}\right)\nonumber \\
\frac{\partial u_i}{\partial t}&=&-\frac{\partial}{\partial r_j}\left(u_i E_{u_j} +
\int d\vv f v_i\frac{\partial E_f}{\partial v_j}\right)\nonumber \\
&&+\frac{\partial}{\partial r_i}\left(-\int d\vv \epsilon +\rho E_{\rho}+sE_s+
u_jE_{u_j}+\int d\vv f E_f\right)\nonumber \\
\frac{\partial f}{\partial t}&=&
-\frac{\partial}{\partial r_i}\left( fE_{u_i}        +f\frac{\partial \eta_f}{\partial v_i}E_s\right)\nonumber \\
&&+\frac{\partial}{\partial v_i}\left(  f\frac{\partial E_{\rho}}{\partial r_i}    +
 f\frac{\partial (E_s\eta_f)}{\partial r_i} +f v_j\frac{\partial E_{u_i}}{\partial r_j}\right)\nonumber \\
&&-\frac{\partial}{\partial r_i}\left(f\frac{\partial E_f}{\partial v_i} \right)+
\frac{\partial}{\partial v_i}\left(f\frac{\partial E_f}{\partial r_i}\right)
\end{eqnarray}
This kinematics-preserving hierarchy  has been already derived  in \cite{GKPhier}.

The disadvantage of the choice of the energy representation (i.e. the disadvantage of the choice of the state variables
$(\rho(\rr),s(\rr), \uu(\rr))$)  is that the transformation    to the entropy representation with the state
variables $(\rho(\rr),\uu(\rr),e(\rr))$, that is more suitable in most applications,  requires an additional assumption. The passage
$(\rho(\rr),\uu(\rr),s(\rr))\rightarrow(\rho(\rr),\uu(\rr),e(\rho(\rr),\uu(\rr),s(\rr))$ is one-to-one only
if $\frac{\partial e(\rr)}{\partial s(\rr)}>0$. The assumption that $\frac{\partial e(\rr)}{\partial s(\rr)}>0$ is in fact a weak form
of the local equilibrium assumption. According to this assumption the quantities $s(\rr)$ and $e(\rr)$ are (local) equilibrium entropy and
(local) energy respectively and consequently $\frac{\partial e(\rr)}{\partial s(\rr)}=T(\rr)>0$ is
the (local) absolute (and thus positive) temperature. We recall that when passing from one representation to another, the gradients change
as follows: $E_{s}= \frac{1}{S_{e}}; E_{\uu}=-\frac{S_{\uu}}{S_{e}}$.

If there is a one-to-one relation between the energy and the entropy representations then the time evolution of the energy
field $e(\rr)$ is governed by
\begin{eqnarray}\label{eqe}
\frac{\partial e}{\partial t}&=&-\frac{\partial}{\partial r_i}\left[\rho E_{\rho}E_{u_i}+s E_sE_{u_i}+
u_jE_{u_j}E_{u_i}\right.\nonumber \\
&&\left.+\int d\vv \left(fE_f E_{u_i} +fE_{\rho}\frac{\partial E_f}{\partial v_i}
+\eta E_s\frac{\partial E_f}{\partial v_i}+fE_f\frac{\partial E_f}{\partial v_i}\right)\right]
\end{eqnarray}

We turn now to  qualitative  properties of solutions of (\ref{eqs11}). A direct consequence of the presence of the  term $\{A,B\}^{(hyd)}$ on the
right hand side of the Poisson bracket (\ref{kh}) (a term that does not involve $f$) the hierarchy  (\ref{eqs11}) is that the hierarchy (\ref{eqs11})
reduces to the standard nondissipative hydrodynamic equations if the energy $E$ is chosen to be independent of $f$,
On the other hand, if $E$ is chosen to depend only on $f$ then (\ref{eqs11}) becomes
the nondissipative one particle kinetic equation.
For a general energy $E$ depending on both the hydrodynamic fields and the one particle distribution function $f$,
the kinematics-preserving  hierarchy (\ref{eqs11}) represents a Hamiltonian extended hydrodynamics in which the
one particle distribution function plays the role of an extra state variable $f$.  The specific physical interpretation of $f$ is
determined by the specification of the energy $E(f,g)$, i.e. by the role that $f$ plays in the forces driving the time evolution.

Having the hierarchy reformulation (\ref{eqs11}) of the kinetic equation (or other hierarchy reformulations discussed in the previous
two sections), how can we use it to make the MaxRent passage to the hydrodynamic equations?  A detail analysis of solutions of (\ref{eqs11}),
in particular an analysis of the onset of irregularities in solutions - see e.g. \cite{Villani}, is expected to lead us to  the upper rate
entropy $\Sigma^{\uparrow}$ generating the dissipation that eventually, by following the dissipative time evolution, eliminates the details
expressed in $f$ and leaves us only with equations governing the time evolution of  hydrodynamic fields. An example  of this type of physical
consideration, but in the context of MaxEnt not MaxRent, is Boltzmann's realization that the binary collisions are responsible for the onset of
irregularities of solutions of the Hamilton one particle kinetic equation and for the emergence  of dissipation in its regularized solutions.
We hope to follow  this route in the future.

In this paper we only note that already the hierarchy reformulation (\ref{eqs11}) is useful in determining the upper rate fundamental
thermodynamic relation. The vector field $J^{\uparrow}(f;\rho,s,\uu)$ is read in the second terms on the right hand side of the equations
governing the time evolution of the hydrodynamic fields. Moreover, the first two lines in the equations governing the time
evolution of $f$ in (\ref{eqs11}) indicate also  $J^*(\rho,s,\uu)$. However, we emphasize,  that the passage from (\ref{eqs11})
to the upper rate fundamental thermodynamic relation and to a proof that solutions to (\ref{eqs11}) modified by supplying it with
the dissipative term approach the upper equilibrium state $\breve{x}$ is left unsolved in this paper.

\subsubsection{Euler kinematics-preserving hierarchy}

All three examples  (\ref{Z2}), (\ref{eqsbb}), and (\ref{eqs11}) of kinematics-preserving hierarchies are hierarchy reformulations of
equations governing the time evolution of distribution functions. In this last  illustration we present kinematics-preserving
 hierarchy of the Euler hydrodynamic equation. The upper level is the level of fluid mechanics, the upper state variable is the momentum
 field $\uu(\rr)$, and the Poisson bracket
\begin{equation}\label{Euler1}
\{A,B\}=\int d\rr u_i\left(\frac{\partial A_{u_i}}{\partial r_j}B_{u_j}-\frac{\partial B_{u_i}}{\partial r_j}A_{u_j}\right)
\end{equation}
expresses mathematically its kinematics.
 The lower level state variables
 \\$\mathbf{W}=(W_1,...,W_n)$ are introduced by
\begin{equation}\label{Euler2}
W_{\alpha i}=\int d\rr w_{\alpha}(\rr)u_i(\rr)
\end{equation}
where $(w_1(\rr),...,w_n(\rr))$ is a given fixed set of $n$ functions $\mathbb{R}^3\rightarrow \mathbb{R}$.

In order to reformulate the Euler equation as a kinematics-preserving hierarchy involving $\mathbf{W}$ (see (\ref{Euler2}))
 as the lower level state variables, we proceed as in the previous sections.
The functions $A$ and $B$ in (\ref{Euler1}) depend on $\uu$ directly and also indirectly through the dependence on $\mathbf{W}$
(that depends on $\uu(\rr)$ - see (\ref{Euler2})).
We thus replace $A_{\uu}$ and $B_{\uu}$  in (\ref{Euler1}) with
$A_{u_i(\rr)}=w_{\alpha}(\rr)A_{W_{\alpha i}}+A_{u_i(\rr)}$ and with the same expression for $B_{\uu}$. This change  transforms (\ref{Euler1})
into
\begin{eqnarray}\label{E1}
\{A,B\}&=& \int d\rr\left[u_iw_{\beta}\frac{\partial w_{\alpha}}{\partial r_j}\left(A_{W_{\alpha i}}B_{W_{\beta j}}-B_{W_{\alpha i}}A_{W_{\beta j}}
\right)\right.\nonumber \\
&&\left.+u_i\frac{\partial w_{\alpha}}{\partial r_j}\left(A_{W_{\alpha i}}B_{u_j}-B_{W_{\alpha i}}A_{u_j}\right)\right.\nonumber \\
&&\left. +u_iw_{\beta}\left(\frac{\partial A_{u_i}}{\partial r_j}B_{W_{\beta j}}-\frac{\partial B_{u_i}}{\partial r_j}A_{W_{\beta j}}\right)\right.
\nonumber \\
&&\left.+u_i\left(\frac{\partial A_{u_i}}{\partial r_j}B_{u_j}-\frac{\partial B_{u_i}}{\partial r_j}A_{u_j}\right)\right]
\end{eqnarray}
In this Poisson bracket (as well as in the Poisson bracket (\ref{Z1})) the upper state variable $\uu$ appears in all its terms. The Poisson brackets
(\ref{BBB}) and (\ref{kh}) involving only the  lower state variables are rather exceptional.

The time evolution equations corresponding to this bracket is the following kinematics-preserving hierarchy
\begin{eqnarray}\label{E2}
\dot{W}_{\alpha i}&=& \int d\rr \left[ \left(u_iw_{\beta}\frac{\partial w_{\alpha}}{\partial r_j}-u_jw_{\alpha}\frac{\partial w_{\beta}}{\partial r_i}
\right)E_{W_{\beta j}}\right.\nonumber \\
&& \left. -u_jw_{\alpha}\frac{\partial E_{u_j}}{\partial r_i}+u_i\frac{\partial w_{\alpha}}{\partial r_j}E_{u_j}\right]\nonumber \\
\frac{\partial u_i}{\partial t}&=&-\frac{\partial}{\partial r_j}\left(u_iE_{u_j}+u_iw_{\beta}E_{W_{\beta j}}\right)-\frac{\partial p}{\partial r_i}
\end{eqnarray}
where  $p=-e+u_jw_{\beta}E_{W_{\beta j}}+u_jE_{u_j} $ and $E(\uu,\WW)=\int d\rr e(\uu,\WW;\rr)$.
The momentum field $\uu(\rr)$ appearing in (\ref{E2}) can be physically interpreted as an average momentum field and $\WW$ as its fine
internal structure.
A possible suitability of this reformulation of the Euler equation for, for example, turbulence modeling or  numerical investigations
   is intended to be explored in a future paper.

\section{Concluding Remarks}

Multiscale thermodynamics is a theory of relations among  levels of investigation of complex systems. It is a theory that sprung  from the
classical equilibrium thermodynamics, Boltzmann's kinetic theory and the Gibbs equilibrium statistical mechanics. A level is
well established if its predictions agree with results of  experimental observations. A level $\mathcal{L}$ is an upper level
vis-\`{a}-vis another level \textit{l} if $\mathcal{L}$ includes more details than \textit{l}.
If both levels $\mathcal{L}$ and \textit{l} are well established then there must exist a way to prepare the systems under investigation
for the level  \textit{l} and the preparation process has to be possible to see as a time evolution on the level $\mathcal{L}$
The  entropy appearing in the vector field governing such time evolution plays on the level $\mathcal{L}$
the role of an ambassador of the lower level \textit{l}. During the time evolution the entropy is maximized
 subjected to certain constraints that, as well as the entropy,  represent the  lower level \textit{l} inside the upper level
 $\mathcal{L}$.
The time evolution in $\mathcal{L}$ leading to \textit{l}
 is a sequence of infinitesimal contact-structure-preserving
  transformations and the whole process of passing from the level $\mathcal{L}$ to the level \textit{l}
 is a reducing Legendre transformation.
Multiscale thermodynamics investigates the chain  $\longrightarrow\mathfrak{L}\longrightarrow \mathcal{L}\longrightarrow l \longrightarrow$,
where $\mathfrak{L}$ is a level that involves more details that both the level $\mathcal{L}$ and \textit{l}. More generally,
the chain is replaced by an oriented graph with levels as its vertices and links, directed toward lower levels, as reductions.

In this paper we first present the main tenets   of the multiscale thermodynamics (in Sections \ref{structure} and \ref{rt}) and
then in Section \ref{realizations}
we show its realizations in the setting of  classical theories like the Boltzmann kinetic theory, Gibbs equilibrium statistical mechanics, and
fluid mechanics.
Dynamic and static  theories on a  wide range of scales become particular realizations of a single abstract theory applicable to externally
and internally unforced and forced complex systems
with no limitations regarding the closeness to equilibrium.
In Section \ref{hierarchy}, we turn the multiscale thermodynamics towards
a new path in hierarchy reformulations of dynamical theories.  Our objective is to formulate hierarchies   that preserve  kinematics.
In both the classical and the newly explored theories the multiscale
thermodynamics inspires novel  insights and viewpoints.
\\

\textbf{Acknowledgemets}: I would like to thank V\'{a}clav Klika, Hans Christian \"{O}ttinger, and Michal Pavelka for inspiring discussions.
\\

\textbf{Conflicts of interest}: The author declares no conflict of interest
\\

\end{document}